\documentclass[pdflatex,sn-mathphys, iicol]{sn-jnl}
\usepackage{color}
\usepackage[capitalise]{cleveref}
\usepackage{physics}
\usepackage[version=4]{mhchem}
\usepackage{supertabular}
\usepackage[uncertainty-mode=separate]{siunitx}

\usepackage{soul}

\AtBeginDocument{\RenewCommandCopy\qty\SI}

\jyear{2026}
\begin{document}

\title[Charge quietness]{Absence of Charge Offset Drift in a Transmon Qubit}

\author[1]{A. Rospars}
\author[1]{H. Hutin}
\author[1]{Y. Seis}
\author[2]{C. Lled\'o}
\author[1]{R. Assouly}
\author[1]{R. Cazali}
\author[1]{R. Dassonneville}
\author[1]{A. Peugeot}
\author[3]{A. Blais}
\author[1]{A. Bienfait}
\author*[1]{B. Huard} 
\email{benjamin.huard@ens-lyon.fr}

\affil[1]{Ecole Normale Sup\'erieure de Lyon,  CNRS, Laboratoire de Physique, F-69342 Lyon, France}
\affil[2]{Departamento de Física, Facultad de Ciencias Físicas y Matemáticas, Universidad de Chile, Santiago 837.0415, Chile}
\affil[3]{Institut Quantique and D\'epartement de Physique,
Universit\'e de Sherbrooke, Sherbrooke J1K 2R1 QC, Canada}

\abstract{Superconducting quantum circuits are sensitive to their electrostatic environment: uncontrolled charges accumulating on the electrodes of a Josephson junction shift the energy levels of a qubit, perturbing its operation and restricting their design. This effect is captured by a single parameter — the charge offset — whose slow, unpredictable drift has proven difficult to eliminate in practice. Here, we report a tantalum-based transmon qubit in which the charge offset remains pinned at zero over nearly three months of measurements, including two thermal cycles, with no observable compromise to the qubit lifetime. This exceptional stability disappears in later cooldowns, indicating a fragile mechanism at play. We attribute it to the inductance of a thin superconducting layer inadvertently formed in parallel with the Josephson junction during fabrication. X-ray surface spectroscopy suggests this layer arises from an incomplete wet-etch of tantalum on sapphire. Deliberately engineering such a layer offers a route to eliminating charge-offset drift in superconducting circuits more broadly.}

\maketitle

Quantum devices are surrounded by uncontrolled fluctuating electric fields. This so-called $1/f^\alpha$ noise is partly produced by trap states or fluctuators in the vicinity of the circuit, fallout from high-energy particle impacts, or by electrical noise in the control lines~\cite{dutta_low-frequency_1981,kogan_electronic_1996,weissman_frac1f_1988,koelle_high-transition-temperature_1999,grasser_noise_2020,wilen_correlated_2021,thorbeck_two-level-system_2023,falci_1f_2024,larson_quasiparticle_2025}. This fluctuating electric environment limits the quantum coherence of charge qubits deriving from the Cooper pair box~\cite{Nakamura1999,Vion2002,bladh_single_2005,metcalfe_measuring_2007}, which led to the design of qubits with reduced sensitivity to charge noise, such as the transmon~\cite{Koch2007}, the fluxonium~\cite{Manucharyan2009} or the inductively shunted transmon (IST)~\cite{hassani_inductively_2023}. The effect of the surrounding electric environment on a Josephson junction is captured by a single parameter: the \emph{charge offset} $n_g$, which represents the number of Cooper pairs electrostatically induced on the island by the environment. The limit on readout power, hence on readout speed, of transmon qubits has been shown to depend on $n_g$~\cite{khezri_measurement-induced_2023,cohen_reminiscence_2023,dumas_measurement-induced_2024,fechant_offset_2025,wang_probing_2025}. Josephson junction arrays can also be sensitive to drifts of the offset charge on the islands of the array, which is a limitation for the operation of some protected qubits~\cite{groszkowski_coherence_2018}. Over the last decade, many experiments have been able to probe the dynamics of $n_g$ in transmons or Cooper pair boxes, and all of them observe $n_g$ drifting with timescales ranging from minutes to hours~\cite{Riste2013,serniak_hot_2018,christensen_anomalous_2019,serniak_direct_2019,Tomonaga2021,wilen_correlated_2021,iaia_phonon_2022,pan_engineering_2022,tennant_low-frequency_2022,wills_spatial_2022,martinez_noise-specific_2023,thorbeck_two-level-system_2023,amin_direct_2024,kamenov_suppression_2024,krause_quasiparticle_2024,liu_observation_2024,larson_quasiparticle_2025,sundelin_real-time_2026,kerschbaum_assessing_2026}.  Here, we report an experiment in which the charge offset $n_g$ of a transmon qubit was measured to be zero in every measurement over the course of nearly three months. We attribute this stable charge offset to the presence of a thin superconducting layer in parallel with the Josephson junction, which is so inductive that it marginally distorts the transmon energy spectrum and relaxation, beyond what can be obtained with spiral inductors~\cite{hassani_inductively_2023}. Surface analysis reveals that wet etching of tantalum on sapphire leaves tantalum traces, which could form such a highly inductive superconducting path to ground.
In later cooldowns, the charge offset started to drift, albeit more slowly than in state-of-the-art charge qubits, indicating the fragility of the structure at the origin of charge stability. The ability to prevent the charge offset from drifting as a by-product of standard qubit fabrication could remove an important constraint in building superconducting circuits.

\begin{figure}[h!]
    \centering
    \includegraphics[width=\columnwidth]{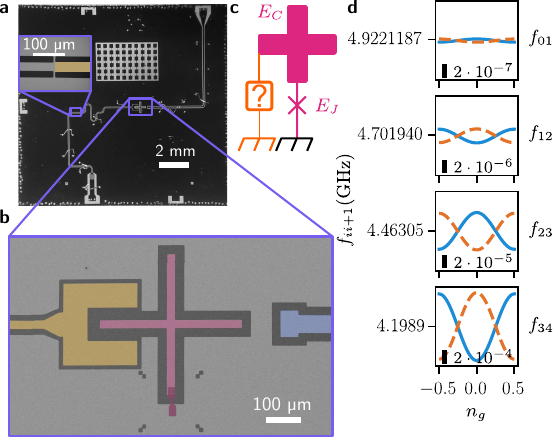}
    \caption{\textbf{Device and charge dispersion.} \textbf{a}, Optical image of the device. The circuit is made of tantalum on sapphire except the Al/AlOx/Al Josepshon junction. Inset: inductive coupling between the readout resonator and its Purcell filter. \textbf{b}, False-colored electron-beam image of the device which includes a transmon island (pink), the junction (dark pink), a readout resonator (yellow) and a charge line (blue). \textbf{c}, Circuit schematic where an inductive or resistive element (orange) stabilizes the charge offset $n_g$. \textbf{d}, Charge dispersion of the first four energy transitions \cite{Chitta2022, Groszkowski2021} of the transmon qubit ($E_J=\SI{16.5}{GHz}$, $E_C=\SI{199}{MHz}$). Blue solid line: transition frequencies $f_{i i+1}$ for even number of quasiparticles $n_\mathrm{qp}$ on the island. Orange dashed line: odd $n_\mathrm{qp}$.}
    \label{fig:1}
\end{figure}

\paragraph*{Observation of the charge offset}
Our device is made of a transmon qubit \cite{Koch2007} with a cross-shaped floating island (pink in \cref{fig:1}), capacitively coupled to a $\lambda/4$ coplanar waveguide (yellow) acting as a readout resonator, and a charge line (blue). The readout resonator is inductively coupled to a Purcell filter (see inset of \cref{fig:1}a). This device was already used in other experiments~\cite{Lledo2023, dassonneville_amplifying_2026}. The complete history of our device is shown in the Supplementary Information \cref{sec:history}, including \texttt{runs 17} to \texttt{25} that are analyzed in this work.

The Hamiltonian of the transmon reads $\hat{H}_\mathrm{tr}=4 E_C(\hat{n}-n_g-n_\mathrm{qp}/2)^2-E_J\cos\hat{\varphi}$, where $E_C=e^2/2C$ is the charge energy, $E_J$ the Josephson energy, $\hat{n}$ the number of Cooper pairs, $\hat{\varphi}$ the phase difference across the junction, $n_\mathrm{qp}$ the number of quasiparticles on the island. In the transmon regime, where $E_J \gg E_C$, the transition frequency $f_{ii+1}$ between the states $\ket{i}$ and $\ket{i+1}$ of the transmon can be written as \cite{Koch2007}:
\begin{equation}
    f_{ii+1}^{n_\mathrm{qp}}(n_g) = \overline{f_{ii+1}}+ (-1)^{n_\mathrm{qp}+i}\frac{\Delta f_{ii+1}}{2}\cos(2\pi n_g)\,,\label{eq:frequency_splitting}
\end{equation}
where $\overline{f_{ii+1}}$ is the mean transition frequency, and $\Delta f_{ii+1}$ the charge dispersion.
The charge dispersion $\Delta f_{ii+1}$ is exponentially suppressed with the ratio $E_J/E_C$, but exponentially increases with the energy level as $(128 E_J/E_C)^{i+1}/(i+1)!$~\cite{Koch2007}.

Our device is deep in the transmon regime, with $E_J/h=\SI{16.5}{GHz}$ and $E_C/h=\SI{199}{MHz}$, so that $E_J/E_C = 83$. Thus, its $0-1$ transition frequency varies by less than 100~Hz when $n_g$ changes, which cannot be resolved given the much larger qubit decoherence rate $\Gamma_2/2\pi\approx3-15~\mathrm{kHz}$. However, the effect of the charge offset on the transmon transition frequency becomes clear when looking at transitions between higher excited states as seen in \cref{fig:1}d.

\begin{figure}[t]
    \centering
    \includegraphics[width=\columnwidth]{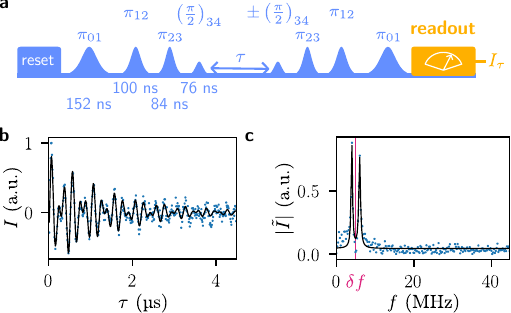}
    \caption{\textbf{Probing the transition frequency $f_{34}$}. \textbf{a}, Pulse sequence applied on the charge line to map Ramsey oscillations between states $\ket{3}$ and $\ket{4}$ onto the $\{\ket{0}, \ket{4}\}$ basis. The $3-4$ pulse is applied at a detuned frequency $\overline{f_{34}}+\delta f=\SI{4.2084}{GHz}$. \textbf{b}, Blue dots: In-phase quadrature as a function of the delay $\tau$ in the Ramsey experiment, averaged over one thousand repetitions. Black line: fit combining two cosine functions ($f_1=\SI{4.0}{MHz}$ and $f_2=\SI{6.0}{MHz}$) with a decay time $T_2^{(34)}=\SI{1.4}{\micro s}$. \textbf{c}, Blue dots: Fourier transform $\lvert\tilde{I}(f)\rvert$ of the same signal. Black line: Fit with a combination of two Lorentzian peaks centered at $f_1$ and $f_2$. Pink vertical line: mean frequency $\delta f=\SI{5.0}{MHz}$ indicating $\overline{f_{34}}=\SI{4.2034}{GHz}$.}
    \label{fig:2}
\end{figure}

To determine the charge offset~\cite{tennant_low-frequency_2022}, we measure the transition frequency $f_{34}$ using the pulse sequence of \cref{fig:2}a. The transmon is prepared in $(\ket{3}+\ket{4})/\sqrt{2}$ and evolves freely for a time $\tau$ before state $\ket{3}$ is mapped onto state $\ket{0}$ and the qubit is read out~\cite{tennant_low-frequency_2022}. The resulting average quadrature $I$ of the readout signal shows Ramsey oscillations between states $\ket{3}$ and $\ket{4}$ (\cref{fig:2}b). The oscillations exponentially decay and exhibit a beating pattern. The two frequencies of the beating correspond to the values of $f_{34}$ for even and odd quasiparticle numbers $n_\mathrm{qp}$. The quasiparticle parity switching rate $\Gamma_{\rm ps}$ varies from run to run between $2.5$ and $\SI{8.0}{kHz}$ (see Methods). Since $\Gamma_{\rm ps}^{-1}$ is shorter than the acquisition time of each experimental point, all the measurements in the figures of the main text are averaged over the two parities of $n_\mathrm{qp}$. The transition frequencies $f^{\textrm{odd}}_{34}$ and $f^{\textrm{even}}_{34}$ can be extracted from the peaks in the Fourier transform of the signal, as shown in \cref{fig:2}c.  By tracking the difference $f^{\textrm{odd}}_{34}-f^{\textrm{even}}_{34}$, one infers the dynamics of the charge offset.

\begin{figure}[t]
    \centering
    \includegraphics[width=\columnwidth]{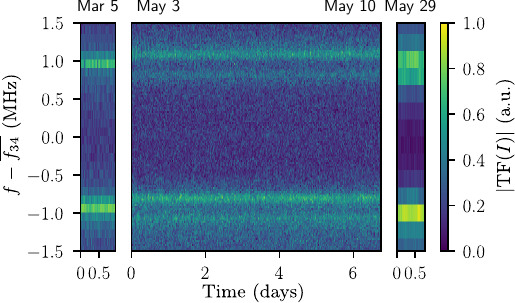}
    \caption{\textbf{Absence of charge offset drift.} Continuous record of the measurement presented in \cref{fig:2}c. The left plot was measured on March 5th, 2024 during \texttt{run 17} every $\SI{17}{s}$, the middle plot was measured from May 3rd to May 10th, 2024 during \texttt{run 17} every $\SI{58}{s}$, and the right plot was measured on May 29th, 2024 during \texttt{run 18} every $\SI{65}{s}$. The frequency splitting is constant, indicating a constant value of the charge offset. In the middle panel, the lines are split into two doublets, which we attribute to a two-level system dispersively coupled to the qubit.}
    \label{fig:3}
\end{figure}

The main result of this work is presented in \cref{fig:3}, where we repeat the Ramsey experiment continuously during three separate time intervals occurring within a period of nearly three months (\texttt{runs 17} and \texttt{18}), during which the dilution refrigerator was thermally cycled twice. In contrast to the usual behavior associated with drift or jumps in the offset charge $n_g$, the transition frequency does not wander on its dispersion curve. Instead, the measured $f_{34}$ frequencies remain stable for up to one week during each measurement interval. These observations indicate that the charge offset remained stable over a period of nearly three months. The linewidth of the frequency peaks is set by the inverse of the maximal wait Ramsey time $\tau$ in the left $(\SI{0.11}{MHz})$ and right $(\SI{0.22}{MHz})$ panels of \cref{fig:3} while it is limited by the decoherence rate of the $3-4$ transition $(\SI{0.08}{MHz})$ for the middle panel.
Note that during the week-long measurement of the middle panel, the frequency peaks come in pairs of doublets. We attribute the splitting into a doublet to the coupling of a two-level system (TLS)~\cite{liu_observation_2024} that we observed in some of the 25 runs of our device (see Supplementary \cref{sec:history} and Fig.~12 in \cite{dassonneville_amplifying_2026}). In an attempt to tune the charge offset, we applied a DC voltage bias on the charge line, but did not observe any change in the transition frequencies. 

\begin{figure}[h!]
    \centering
    \includegraphics[width=\columnwidth]{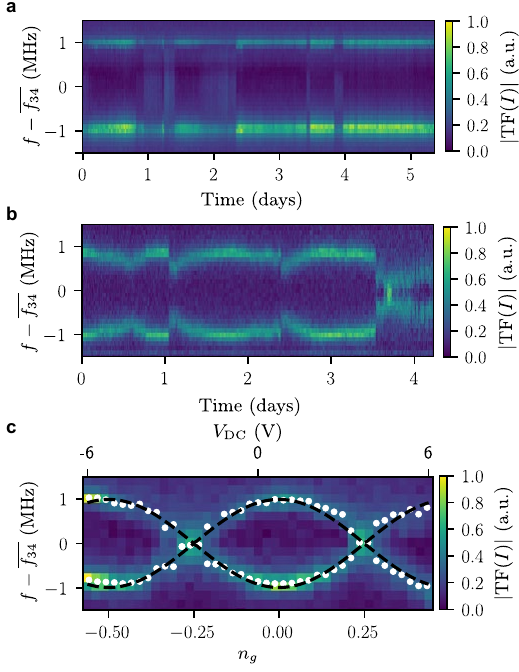}
    \caption{\textbf{Evolution of the charge offset in later runs.} \textbf{a}, Record of $f_{34}$ using the protocol presented in \cref{fig:2}c over more than five days during \texttt{run 21}, every $\SI{231}{s}$. Color: Fourier transform of Ramsey oscillations of the $3-4$ transition. \textbf{b}, Record of $f_{34}$ obtained during \texttt{run 25}, every $\SI{16}{s}$. \textbf{c}, $3-4$ transition frequencies as a function of applied voltage $V_{\rm DC}$ on the charge line (referred to the source) and corresponding value of charge offset $n_g$ during \texttt{run 22}. The whole measurement takes $\SI{38}{s}$, shorter than the charge drift timescale. White dots: fitted transition frequencies. Black dashed line: sinusoidal fit of the white dots $f_{34}^\pm=\overline{f_{34}}\pm \frac{\Delta f_{34}}{2}\cos(2\pi n_g)$. Here $\Delta f_{34} = \SI{2}{MHz}$ and $n_g = n_g^0+\frac{V_{\rm DC}}{12~{\rm V}}$.}
    \label{fig:4}
\end{figure}

\paragraph*{Charge offset becomes unstable}
Remarkably, after opening the sample holder for visual inspection and improving the shielding (see Supplementary \cref{sec:history}), the Ramsey sequence measurement revealed variations of the $f_{34}$ transition frequency starting with \texttt{run 20} (Figs.~\ref{fig:4}a, \ref{fig:4}b, \ref{fig:sup_wire bond}, and \ref{fig:sup_23}). The charge offset could then be tuned by applying a DC voltage $V_\mathrm{DC}$ to the charge line, and the transition frequencies $f^{\textrm{odd}}_{34}$ and $f^{\textrm{even}}_{34}$ depend on $V_\mathrm{DC}$ as expected from \cref{eq:frequency_splitting}. This allows us to determine the value of the charge dispersion $\Delta f_{34}=\SI{2.0}{MHz}$ (\texttt{run 22} in Fig.~\ref{fig:4}c). It matches the frequency splitting observed in \cref{fig:3}, confirming that the charge offset was pinned at $n_g=0$ during \texttt{runs 17} and \texttt{18}. Using the dependence of the transition frequency $f_{34}$ on $n_g$, the measured linewidth of the frequency peaks translates directly into a bound on charge offset fluctuations. We find that $n_g$ remained at $0 \pm 0.06$ throughout \texttt{runs 17} and \texttt{18}.

In \texttt{run 21}, while the charge offset is no longer steady for months, its evolution is characterized by day-long periods at zero charge offset $n_g=0$, and other time intervals during which the transition frequency $f_{34}$ cannot be resolved as the coherence time dramatically drops (see \cref{fig:4}a). In later runs (such as \texttt{run 25} in \cref{fig:4}b), the charge offset is very noisy during some periods and quieter during others. In contrast to \texttt{run 21}, the charge offset drifts and jumps during the quieter times, which is a standard behavior reported in many previous works~\cite{Riste2013,serniak_hot_2018,christensen_anomalous_2019,serniak_direct_2019,Tomonaga2021,wilen_correlated_2021,iaia_phonon_2022,pan_engineering_2022,tennant_low-frequency_2022,wills_spatial_2022,martinez_noise-specific_2023,thorbeck_two-level-system_2023,amin_direct_2024,kamenov_suppression_2024,krause_quasiparticle_2024,larson_quasiparticle_2025,liu_observation_2024,sundelin_real-time_2026,kerschbaum_assessing_2026} albeit more slowly here. The noise spectrum of the charge offset follows a $1/f^2$ dependence on frequency, similar to what was reported in Refs.~\cite{christensen_anomalous_2019,tennant_low-frequency_2022} but with a smaller prefactor (see Methods). It thus seems that the mechanism responsible for the charge offset cancellation was intermittently present in \texttt{run 21} and completely absent from \texttt{run 22} onward.

\paragraph*{Origin of the charge stability}

We conducted several investigations to elucidate the observed charge stability (see Supplementary Information). Our main hypothesis is that an inductive path exists in parallel with the Josephson junction, which allows electrostatically induced Cooper pairs to equilibrate between the cross-shaped island and the ground plane (see \cref{fig:1}c). The geometry and topology of this inductive path is unknown. Assuming that we can effectively model the circuit as an inductively shunted transmon~\cite{koch_charging_2009,hassani_inductively_2023}, the Hamiltonian $\hat{H}_\mathrm{tr}$ gains an extra term $\frac{E_L}{2}(\hat{\varphi}+\varphi_\mathrm{ext})^2$, where $\varphi_\mathrm{ext}$ is the reduced flux in the loop, and $E_L=\varphi_0^2/L$ is the inductive energy associated with the inductance $L$ and $\varphi_0=\hbar/(2e)$ is the reduced flux quantum. 
Although no external magnetic field was applied during the runs where $n_g$ was pinned to zero, it is still possible to set a lower bound for $L$. Considering the small change in transition frequencies when the inductance disappeared in the latest runs (\cref{fig:freqchange}), we find $E_L/h<0.03~\mathrm{GHz}$ (see Supplementary Information \cref{sec:inductancebound}), or equivalently $L>20~\mathrm{\mu H}$. This would be a particularly large inductance that goes beyond the carefully engineered inductances of disordered superconductors~\cite{khorramshahi_high-impedance_2025}, Josephson junction arrays~\cite{pechenezhskiy_superconducting_2020,junger_implementation_2025}, or spiral inductors~\cite{hassani_inductively_2023}. The observation of charge-parity jumps in the transition frequency may appear incompatible with the presence of an inductive path to ground~\cite{Manucharyan2009, koch_charging_2009}, which would normally allow the island charge to equilibrate. However, if the path to ground is made of a superconductor with a larger gap than the Al junction leads (such as tantalum), low-energy quasiparticles in the Al cannot diffuse into it. Consequently, the inductive path to ground does not suppress charge-parity switching.

An alternative explanation is that a resistive path exists in parallel with the Josephson junction. The value of the resistance $R$ needs to be compatible with the measured transmon lifetime $T_1\approx 21-24~\mu\mathrm{s}$. Knowing the transmon capacitance $C=0.1~\mathrm{pF}$ from $E_C$, the resulting $RC$ time thus needs to be greater than $T_1$. This sets a lower bound on $R$ well above 0.2~G$\Omega$. Such a resistance is well above the resistance quantum and would be difficult to achieve with known resistive materials spanning the 100~$\mu$m gap between the transmon island and the ground plane, especially without introducing dielectric losses.

To determine the possible origin of the inductive or resistive element in parallel with the junction, we turned to surface analysis. Energy dispersive X-ray (EDX) spectroscopy reveals the unexpected presence of tantalum between the transmon island and the ground plane. Complementary X-ray photoelectron spectroscopy (XPS) was performed in two different laboratories to characterize the first few nanometers of the etched regions in other parts of the device. Both measurements found large fractions of Ta and Ta$_2$O$_5$ where tantalum was supposed to be etched away (see Methods). During the fabrication of the device, wet etching was performed by dipping the tantalum covered sapphire chip for $\SI{17}{s}$ in Transene Tantalum Etchant 111 at room temperature after optical lithography (see Supplementary \cref{sec:fabrication}). Other XPS analyses were performed on devices that were etched for up to $\SI{60}{s}$, a time long enough to also etch the tantalum regions initially covered by optical resist, and tantalum was still found in the etched regions. We conclude that this wet etching of sputtered tantalum on sapphire is incomplete. It is possible that some etch resistant interfacial layer remains, possibly related to the recently observed intermixing layer between tantalum and sapphire~\cite{mcfadden_interface-sensitive_2025,anbalagan_revealing_2025,olszewski_krypton-sputtered_2026}. We thus conjecture that, under some tantalum sputtering and etching conditions, the remaining tantalum rich layer provides a highly inductive path between the island and the ground plane.

\paragraph*{Conclusion}

We have demonstrated that the charge offset of a transmon-like qubit can remain pinned to zero for months, while preserving lifetimes of several tens of microseconds. Unlike circuits insensitive to charge offset such as the fluxonium~\cite{Manucharyan2009}, the device retains the signature of a transmon with transition frequencies that jump between two values as the charge parity switches. Surface analysis showed that an incomplete tantalum etch, likely coming from an intermixing layer between tantalum and sapphire~\cite{mcfadden_interface-sensitive_2025,anbalagan_revealing_2025,olszewski_krypton-sputtered_2026}, can leave a tantalum rich layer between the island and the ground. This film could form an inductive or resistive path between the island and the ground plane, which removes the frequencies' dependence on charge offset. In both scenarios, the inductance ($L>20~\mathrm{\mu H}$) or the resistance ($R\gg0.2~\mathrm{G\Omega}$) are much larger than what can be reached with standard materials, making this an unusual and potentially unique regime of inductive or resistive shunting. 

The gradual return to standard charge-offset behavior over successive cooldowns suggests that the structure responsible for charge stability is fragile. Yet, even without complete protection against charge offset drifts, $n_g$ evolves more slowly than other works (\cref{fig:sup_SND}). If such an inductive layer could be engineered reproducibly--for instance by tuning deposition conditions or surface treatment--it could offer a practical route to eliminating charge-offset drift. We speculate that anomalous charge-offset dynamics reported in previous tantalum-based transmon experiments~\cite{tennant_low-frequency_2022} may share the same origin as the stability we observe here, suggesting that the an inductive or resistive path to ground may be more common than one would expect. Understanding and controlling the residual tantalum layer therefore appears as a promising direction not only for improving qubit stability, but for engineering the electrostatic environment of superconducting quantum circuits more generally.

\subsection*{Acknowledgments}
This research was supported by the ANR OCTAVES (ANR-21-CE47-0007), and GASP (Contract No. W911-NF23-10093) programs. C.L. was supported by the National Agency for Research and Development (ANID) through FONDECYT Postdoctoral Grant No.~3250130. We acknowledge IARPA and Lincoln Labs for providing a Josephson Traveling-Wave Parametric Amplifier. We thank Christian Perruchot, Philippe Decorse from the ITODYS XPS facility (Université Paris Cité, CNRS UMR 7086, PARIS, France), and Corinne Lagrost and Jules Galipaud from the ASPHERYX platform (ScanMAT, UAR 2025, University of Rennes-CNRS; CPER MAT \& Trans 2021-2027) for the XPS spectroscopies. We also thank Kassandra Gerard and Cendrine Moskalenko from LPENSL for performing and analyzing the AFM scans. We are grateful to Patrice Bertet, Landry Bretheau, Andrew Cleland, Michel Devoret, Timothy Duty, Mathieu Fechant, Manuel Houzet, Julia Meyer, William Oliver, Ioan Pop, Hugues Pothier, Nicolas Roch, Max Schaefer, David Schuster, Irfan Siddiqi and Simon Zihlmann for fruitful discussions.

\subsection*{Author contributions}
A.R., H.H. and Y.S. performed the experiments and analyzed the data. R.A., R.C. and R.D. fabricated the sample. A.R., H.H., Y.S., C.L., A.Bi. and B.H. analyzed and discussed the results. A.P. and A.Bl. provided additional support for the experiments and analyses. A.R. and B.H.  wrote the manuscript with comments from all the authors. B.H. supervised the project.

\section*{Methods}

\begin{figure}[h!]
    \centering
    \includegraphics[width=\columnwidth]{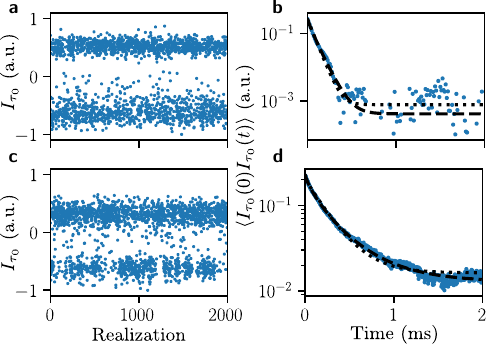}
    \caption{\textbf{Parity switching}. \textbf{a}, Signal $I_{\tau_0}$ for $2,000$ realizations in the absence of charge drift (situation of \cref{fig:3}, \texttt{run 17}). The signal alternates between two values, corresponding to the even and odd parities of the charge offset. The measurement is repeated every $\SI{12}{\micro s}$ on average. \textbf{b}, Autocorrelation function $\langle I_{\tau_0}(0)I_{\tau_0}(t)\rangle$ extracted from more than $10^6$ realizations of $I_{\tau_0}$ during \texttt{run 17}. The black dotted and dashed correspond respectively to an exponential and a stretched exponential fit. \textbf{c}, Signal $I_{\tau_0}$ for $2,000$ realizations in the presence of charge drift (\texttt{run 24}). The measurement is repeated every $\SI{3}{\micro s}$ on average. \textbf{d}, Autocorrelation function $\langle I_{\tau_0}(0)I_{\tau_0}(t)\rangle$ for \texttt{run 24}. Black: fits as in b.}
    \label{fig:parity-switching}
\end{figure}

\paragraph{Parity switching rate}\label{sec:parity-switching}
To measure the parity switching rate, we perform the Ramsey experiment on the $3-4$ transition presented in \cref{fig:2}. A pulse is sent on resonance with one of the two peaks, and the time delay is fixed to $\tau_0=\frac{1}{2\Delta f_{34}}$. For each realization, we save the in-phase quadrature $I_{\tau_0}$ of the demodulated signal as seen on \cref{fig:parity-switching}a and c. The measured data alternate between two values, corresponding to the even and odd parities of the quasiparticle number $n_\mathrm{qp}$.

The parity switching rate $\Gamma_{\rm ps}$ is extracted from the decay rate of the autocorrelation of $I_{\tau_0}$ \cite{Riste2013}, as shown in \cref{fig:parity-switching}b and d. The decay of the autocorrelation function $\langle I_{\tau_0}(0)I_{\tau_0}(t)\rangle$ is not exactly exponential (dashed black line) indicating non-Markovian effects \cite{Nho2025, Kurilovich2025} but can be well fitted by a stretched exponential of the form $Ae^{-(2t\Gamma_{\rm ps})^\beta}+B$. During \texttt{run 17} without charge drift, we extract a parity switching rate $\Gamma_{\rm ps} = \SI{8.0}{kHz}$ and $\beta=0.89$, as shown in \cref{fig:parity-switching}b. In the presence of charge drift, we extract a parity switching rate $\Gamma_{\rm ps} = \SI{2.5}{kHz}$ and $\beta=0.75$, as shown in \cref{fig:parity-switching}d. We note that fitting the autocorrelation function to a regular exponential decay ($\beta=1$) leads to similar rates $\Gamma_{\rm ps}^\mathrm{markov} = \SI{7.5}{kHz}$ and $\Gamma_{\rm ps}^\mathrm{markov}= \SI{2.0}{kHz}$, respectively. These relatively high rates compared to the state-of-the-art (below Hz) are likely due to the sample holder which has many sub-mm gaps around its connectors, offering a poor protection against infrared radiation. 

In addition to determining the parity switching rate, these single-shot measurements provide the imbalance between the times spent with each parity. Using over one million realizations, we determine the even/odd parity imbalance to be $51\%-49\%\pm0.1\%$ (\texttt{run 17} without charge offset drifts) and  $54\%-46\%\pm0.1\%$ (\texttt{run 24} with charge offset drifts).

\paragraph{Spectral noise density}\label{sec:SND}
In order to quantify the evolution of the charge offset drifts, we repeated the experiment presented in \cref{fig:4}c. As discussed in \cite{christensen_anomalous_2019}, it is important to perform the full spectroscopy as function of $V_{\rm DC}$ instead of only inferring $n_g$ from the frequency splitting, which would lead to underestimating the noise. We realized such a measurement during \texttt{run 22}, where we were able to apply a DC voltage on the drive line. This was not possible in later runs when the drive line was connected to the ground through a wire bond (see corresponding investigation in Supplementary Information).

\begin{figure}[t]
    \centering
    \includegraphics[width=\columnwidth]{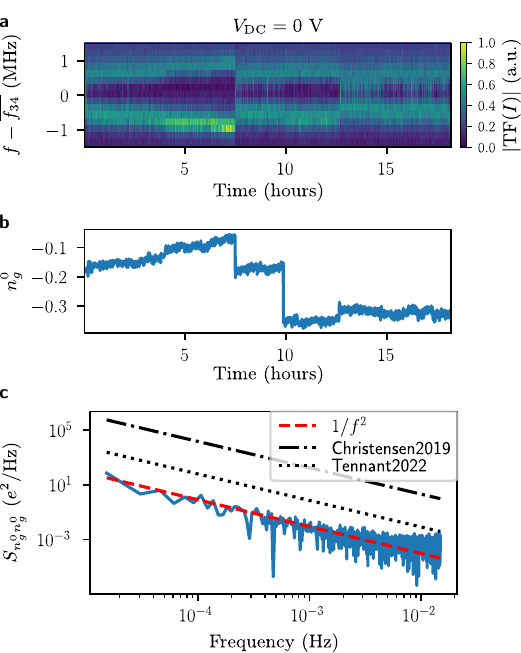}
    \caption{\textbf{Noise spectral density}. \textbf{a}, Continuous record of the measurement of $f_{34}$ presented in \cref{fig:2}c ($V_{\rm DC}=\SI{0}{V}$) over more than $18$ hours starting September 1, 2024 during \texttt{run 22}. \textbf{b}, Evolution of the charge offset found by fitting the full spectroscopy as function of $V_{\rm DC}$ during the same measurement as in a. The repetition time is $\SI{23}{s}$. \textbf{c}, Blue curve: spectral noise density of the charge offset presented in b. Black lines: comparison with Refs. \cite{christensen_anomalous_2019, tennant_low-frequency_2022}. Red line: fitting of the noise spectrum with a $1/f^2$ power spectrum.}
    \label{fig:sup_SND}
\end{figure}

By repeating the experiment presented in \cref{fig:4}c for different iterations with a repetition time of $\SI{23}{s}$, one can fit the Fourier spectrum by two Lorentzians to extract the two frequencies $f^\pm_{34}(V_{\rm DC},t)$. A cut of the data at $V_{\rm DC}=\SI{0}{V}$ is shown in \cref{fig:sup_SND}a. Note that in this run, the charge offset drifts on timescales much faster than in other runs (\cref{fig:4} or \cref{fig:sup_wire bond}).

Then, we fit all the frequencies simultaneously by $f_{34}^\pm (V_{\rm DC},t) = \overline{f_{34}}\pm \frac{\Delta f_{34}}{2}\cos\left(2\pi\left(n_g^0(t)+\frac{V_{\rm DC}}{V_{\rm DC}^{\rm period}}\right)\right)$, where the mean frequency $\overline{f_{34}}$, the frequency splitting $\Delta f_{34}$ and the conversion factor between voltage and charge offset $V_{\rm DC}^{\rm period}$ do not depend on time. The evolution of $n_g^0(t)$ is shown in \cref{fig:sup_SND}b.

From the evolution of the charge offset as function of time, one can find the noise spectrum $S_{n_g^0n_g^0}$. The spectrum can be well fitted by a power law with exponent $-2$ and $S_{n_g^0n_g^0}(\SI{0.1}{mHz}) = \SI{0.8}{e^2/Hz}$. This prefactor is smaller than the one reported in similar experiments \cite{christensen_anomalous_2019, tennant_low-frequency_2022} whereas it was obtained in the run presenting the largest charge offset drifts out of all runs. This suggests that, even when suppressed, the phenomenon at the origin of the charge stability in \texttt{runs 17} and \texttt{18} still offers some level of protection against charge offset drift.

\paragraph{Energy dispersive X-ray spectroscopy}\label{sec:EDX}
We perform energy dispersive X-ray spectroscopy (EDX) with a JEOL JSM-IT800 scanning electron microscope (SEM) to determine the elemental composition of the regions where tantalum has been wet-etched and compare it to the tantalum ground plane. These analyses were all performed after the last run.

\begin{figure}[h!]
    \centering
    \includegraphics[width=\columnwidth]{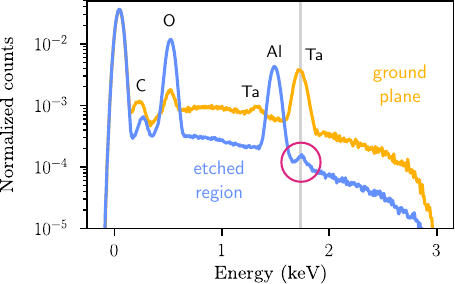}
    \caption{\textbf{EDX analysis of the device with a $\SI{3}{keV}$ electron beam}. The yellow curve is a reference spectrum obtained on the tantalum ground plane. The blue curve is a spectrum measured on a region where tantalum has been wet-etched. This signal is integrated for $\SI{6000}{s}$. The vertical line and circle highlight the peak at $\SI{1.71}{keV}$ corresponding to the $M\alpha$ spectral line of tantalum.}
    \label{fig:sup_edx}
\end{figure}

A typical EDX spectrum of the etched region is shown in \cref{fig:sup_edx} for site I1 (location shown in Fig.~\ref{fig:sup_xps}). We identify four peaks. The oxygen $K\alpha = \SI{0.525}{keV}$  and aluminum $K\alpha = \SI{1.486}{keV}$ emission lines are attributed to the sapphire substrate $\ce{Al2O3}$. The carbon $K\alpha = \SI{0.277}{keV}$ emission line is attributed to surface contamination. Finally, the tantalum $M\alpha = \SI{1.709}{keV}$ emission line is less pronounced but clearly visible despite the wet etching.  The yellow curve is a reference spectrum measured on the tantalum ground plane. We observe the peak of oxygen and two peaks of tantalum, as well as the carbon peak. On the etched regions, the average atomic percentages of the elements (excluding carbon) are fitted to be $x_{\ce{Ta}} = 0.33\pm 0.03 \%$ for tantalum, $x_{\ce{Al}} =39.54 \pm 0.17 \%$ for aluminum, and $x_{\ce{O}} = 60.13 \%$ for oxygen. Note that the peak of tantalum at $\SI{1.335}{keV}$ was only observed on the ground plane, but it is known to vanish for thin tantalum films \cite{coatings11101206}.

Owing to the interaction between incoming electrons and the device materials, the X-ray signal is emitted from a pear shaped region below the surface. For the $\SI{3}{keV}$  electron beam incident on sapphire, the signal integration volume extends laterally and in depth on the order of $\SI{100}{nm}$. Therefore, while traces of tantalum are clear, it seems impossible to determine the exact thickness of the tantalum-rich layer from these EDX measurements. Performing EDX spectroscopy at various positions on etched regions of the device, including near the transmon electrode, always shows a tantalum peak.

\paragraph{X-ray photoelectron spectroscopy}\label{sec:XPS}

\begin{table}[]
    \centering
    \begin{tabular}{|c|c|c|c|c|}
         \hline
         Position&$\ce{Ta/C}$&$\ce{Al/C}$&$\ce{Ta/Ta2O5}$&Comment\\
         \hline
         \textbf{I0}&\textbf{0.3}&\textbf{0}&\textbf{0.6}&\\
         I2&0.2&1.0&$\approx$0.4&\\
         I3&0.2&1.0&$\approx$0.4&device\\
         I4&0.2&1.4&$\approx$0.4&of main\\
         \textbf{A0}&\textbf{0.2}&\textbf{0}&\textbf{0.4}&text\\
         A1&0.1&0.8&0.3&\\
         A2&0.2&1.5&0.3&wet\\
         A3&0.2&0.7&0.3&etched\\
         A4&0.1&3.5&0.2&17~s\\
         \textbf{A5}&\textbf{0.9}&\textbf{0.3}&\textbf{0.3}&\\
         \textbf{A6}&\textbf{3.3}&\textbf{0.6}&\textbf{0.4}&\\
         \textbf{A7}&\textbf{1.8}&\textbf{0}&\textbf{0.4}&\\
         \hline
         \textbf{J0}&\textbf{0.6}&\textbf{0}&$\approx$\textbf{0.6}&wet\\
         J1&0.1&0.7&$\approx$0.5&etched\\
         J2&0.7&0&$\approx$0.5&40~s\\
         J3&0.1&0.8&$\approx$0.5&\\
         \hline
         \textbf{K0}&\textbf{0.6}&\textbf{0}&$\approx$\textbf{1.4}&wet\\
         K1&0.1&0.8&$\approx$0.9&etched\\
         K2&0.6&0.9&$\approx$0.9&60~s\\
         K3&0.1&0.7&$\approx$1.1&\\
         \hline
         \textbf{L0}&\textbf{0.4}&\textbf{0.1}&&\\
         L1&0.4&0.1&$\approx$\textbf{0.4}&HMDS\\
         L2&0&0.7&&before\\
         L3&0.05&0.7&$\approx$0.4&wet\\
         L4&0&0.7&&etching\\
         L5&0&0.7&&17~s\\
         L6&0&0.7&&\\
         \hline
    \end{tabular}
    \caption{\textbf{Molecular fraction of various elements a few nm below the surface.} They are extracted from the XPS peaks at the positions defined in \cref{fig:sup_xps}. The numbers indicate the ratio between molecular content of Ta versus C, Al versus C, and Ta versus \ce{Ta2O5}. Bold characters indicate sites located on the Ta ground plane.}
    \label{tab:composition}
\end{table}

In order to improve the depth resolution of the surface analysis, X-ray photoelectron spectroscopy (XPS) was also performed. As a consistency check, two different laboratories realized the spectroscopies on instruments with different lateral resolutions. Most (I, J, K, and L sites on \cref{fig:sup_xps}) were performed with a Thermo Fischer K-Alpha (\SI{1486.6}{eV}) XPS instrument at ITODYS lab, Universit\'e Paris-Cit\'e with a spot size of $\SI{70}{\micro m}$ (sites I and L) or $\SI{30}{\micro m}$ (sites J and K). Some spectroscopies (sites A in \cref{fig:sup_xps}) were performed on a Thermo Fischer Nexsa G2 instrument ($K\alpha-\ce{Al}$, $\SI{1486.6}{eV}$) at ASPHERYX lab, Université de Rennes with a spot size of $\SI{10}{\micro m}$. Given the spot size and the uncertainty in the localization of the beam, it is impossible to perform relevant spectroscopy around the transmon electrode because of the possibility that the signal partly comes from the tantalum ground plane or electrodes. The analysis was thus realized close to test structures, a few mm away from the transmon electrode, where the etched regions are large enough to avoid any parasitic contribution of the tantalum pads. 
\begin{figure}[h!]
    \centering
    \includegraphics[width=\columnwidth]{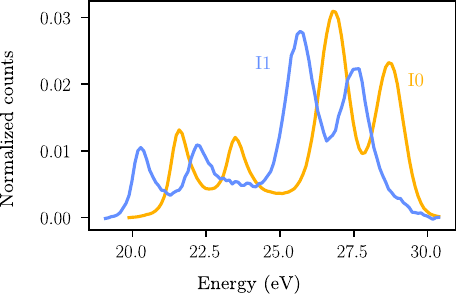}
    \caption{\textbf{XPS high resolution spectra at positions I0 and I1.} Both traces (yellow for I0 and blue for I1) look alike apart from an offset on the energy axis attributed to charging effects. Referred to the blue trace, the peaks around $20$ and $\SI{23}{eV}$ correspond to tantalum, and the peaks around $26$ and $\SI{28}{eV}$ to tantalum oxyde ($\ce{Ta2O5}$).}
    \label{fig:sup_xps_spectra}
\end{figure}

The XPS spectra provide information about the chemical composition of the first few nanometers below the surface of the device. The key features are the presence of an aluminum peak around $\SI{74.5}{eV}$ characteristic of sapphire and two doublets around $20$ and $\SI{23}{eV}$ and around $26$ and $\SI{28}{eV}$ characteristic of tantalum and tantalum oxide (\ce{Ta2O5}) respectively (\cref{fig:sup_xps_spectra}). Fitting these peaks gives a value for the atomic fraction of each chemical entity. In order to compare different sites and different devices, we use the atomic percentage of carbon as a reference. The results of the analysis are presented in \cref{tab:composition}. Measurement series I and A are performed on the device presented in the main text, which was etched for nominally 17~s.
The lines in bold in \cref{tab:composition} indicate sites located in the ground plane. The position of each spectroscopy site is shown in \cref{fig:sup_xps}.

\begin{figure}[t]
    \centering
    \includegraphics[width=\columnwidth]{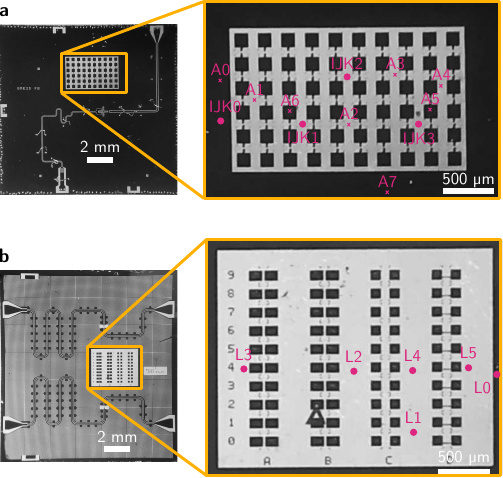}
    \caption{\textbf{Sites investigated by XPS.} \textbf{a}, Optical image of the device presented in the main text (metal is dark and etched tantalum regions are bright) and location of the sites A and I. The same picture is used to locate the sites J and K on the  two similar devices that were wet-etched for a longer duration. \textbf{b}, Another device wet-etched after HMDS coating. Because of the poor lateral resolution of the XPS analysis, the locations were chosen around test junctions and not in the gap around the transmon.}
    \label{fig:sup_xps}
\end{figure}

As expected, in \cref{tab:composition}, spectroscopy of the tantalum ground planes shows a large tantalum content, with no or small signs of aluminum traces. However, in most regions where tantalum has been etched away, the two characteristic doublets of tantalum and tantalum oxide appear, as shown in blue in \cref{fig:sup_xps_spectra}. These doublets reveal the presence of tantalum on the sapphire substrate in the etched regions. The fitted molecular fraction of tantalum and aluminum can be compared. However the ratios are found to strongly depend on position and it is hard to make a quantitative determination of the tantalum layer thickness or length scales from the XPS measurement. 

Suspecting a possible issue with the positioning of the beam, which may lead to unwanted overlap between the etched and pristine tantalum regions, two labs independently performed the analysis as explained above. Both consistently found signatures of tantalum in the etched regions.

Suspecting insufficient etching time with nominally $\SI{17}{s}$, we decided to realize two new devices with the same design but where the wet etching time of tantalum was $\SI{40}{s}$ (series J) and $\SI{60}{s}$ (series K). The tantalum peaks are also visible with similar molecular fraction as the short etching time. This incomplete wet etching could indicate tantalum redeposition during the etching process, or that tantalum has migrated inside the sapphire and is somehow preventing complete wet etching.

Measurement series L was performed on another device that was etched with the same wet etching technique and duration as the device of the main text (see Supplementary \cref{sec:fabrication}), the only difference being the application of a thin layer of hexamethyldisilazane (HMDS) before the spincoating of the optical resist. Surprisingly, no tantalum was detected at most sites. We note that the sapphire wafers used as a substrate are not the same for the device in the main text (A,I) and for the other devices (J,K,L). Likewise, the tantalum sputtering was not done in the same batch.

Finally, we determine the molecular ratio between tantalum and its oxide using the relative heights of the doublets (\cref{tab:composition}). This ratio shows spatial variation and a possible trend towards relatively less oxide for longer etching times, but definite presence. 

\newpage
\renewcommand{\thefigure}{S\arabic{figure}}
\renewcommand{\thetable}{S\arabic{table}}

\section*{Supplementary Information}

\section{Device fabrication}
\label{sec:fabrication}

The device described in the main text is fabricated on a $\SI{430(25)}{\micro m}$ C-plane (001) sapphire wafer, double-side EPI-polished. The wafer is cleaned in a Piranha bath and oxygen-ashed before a $\SI{200}{nm}$-thick tantalum film is sputtered at about $\SI{500}{\celsius}$ by STAR Cryoelectronics (Santa Fe, USA). An optical lithography step is then performed with a Smart Print UV. The tantalum is wet-etched for a nominal duration of $\SI{17}{s}$ in Transene Tantalum Etchant 111 at room temperature. The reaction is stopped by dipping the sample for one minute in deionized water, followed by blow-drying with a nitrogen gun. Note that the resist mask is observed to lift off at the end of the wet etching. We note that the documentation by Transene company indicates an etch rate of $3-\SI{4}{nm\cdot s^{-1}}$ at $\SI{25}{\celsius}$. However, our profilometric measurement shows that the actual etch rate is much faster than this value.

The Josephson junction is fabricated by electron lithography and is made of an Al/AlOx/Al trilayer. It is evaporated at two angles $\pm \ang{35}$ following the Dolan bridge technique. A detailed recipe can be found in chapter 8 of Ref.~\cite{assouly:tel-04057646}. 

For the device hosting locations denoted by L only, right before spin coating the optical resist, the sample is placed inside a desiccator next to a beaker containing HMDS for $\SI{210}{s}$ at $\SI{96}{kPa}$ below atmospheric pressure.

\section{Surface imaging}

\subsection{Scanning electron microscope imaging and atomic force microscopy}\label{sec:sample}\label{sec:afm}

\begin{figure}[t]
    \centering
\includegraphics[width=\columnwidth]{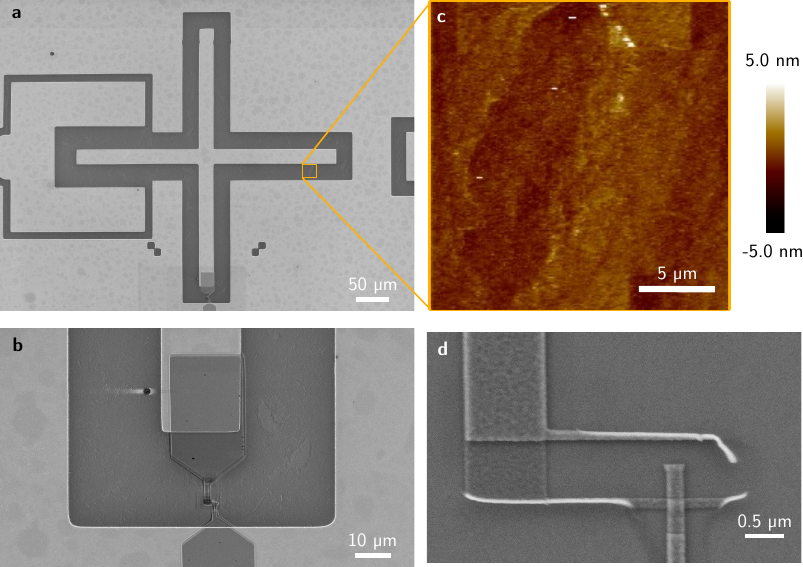}
    \caption{\textbf{Images of the transmon.} \textbf{a}, Electronic image of the cross-shape electrode of the transmon. \textbf{b}, Zoom on the bottom of the cross where the Josephson junction is located. \textbf{c}, AFM image of the orange region indicated in a. The spots seen on the electronic image correlate with holes a few nanometers deep. \textbf{d}, Zoom on the Josephson junction.}
    \label{fig:sup_sample}
\end{figure}

Inspecting the device with a scanning electron microscope, we observed unexpected features between the transmon island and the surrounding ground plane (dark gray regions in \cref{fig:sup_sample}a and b). 
Atomic Force Microscopy (AFM) revealed that these features are about one nanometer deep holes (\cref{fig:sup_sample}c). Their origin is unknown, but they do not provide a clear evidence for an inductive path between the transmon island and the ground plane.

The Josephson junction is shown in \cref{fig:sup_sample}d. Its size is $\SI{130}{nm}$ by $\SI{225}{nm}$. We note that one lead of the junction is partially lifted from the surface. However, no clear inductive path appears in parallel to the junction.

\subsection{Profilometry measurements}\label{sec:dektak}
\begin{figure}[h!]
    \centering
    \includegraphics[width=\columnwidth]{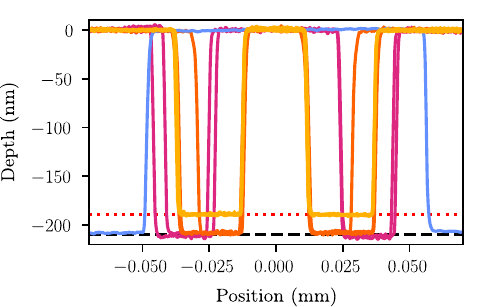}
    \caption{\textbf{Measured profiles of device structures revealing the step height between pristine and etched tantalum regions.} Yellow: profile of the device presented in the main text. Orange: same design etched $\SI{40}{s}$ (series J). Magenta: device presented on \cref{fig:sup_xps}b (series L). Blue: device from the same wafer, hence same sputtered tantalum layer as for the device presented in the main text.
    The device of the main text has a step size of $\SI{189(2)}{nm}$ (red dotted line), whereas the other ones have a step size of $\SI{211(3)}{nm}$ (black dashed line).
    }
    \label{fig:sup_dektak}
\end{figure}
\begin{figure*}[!h]
    \centering
    \includegraphics[width=1\linewidth]{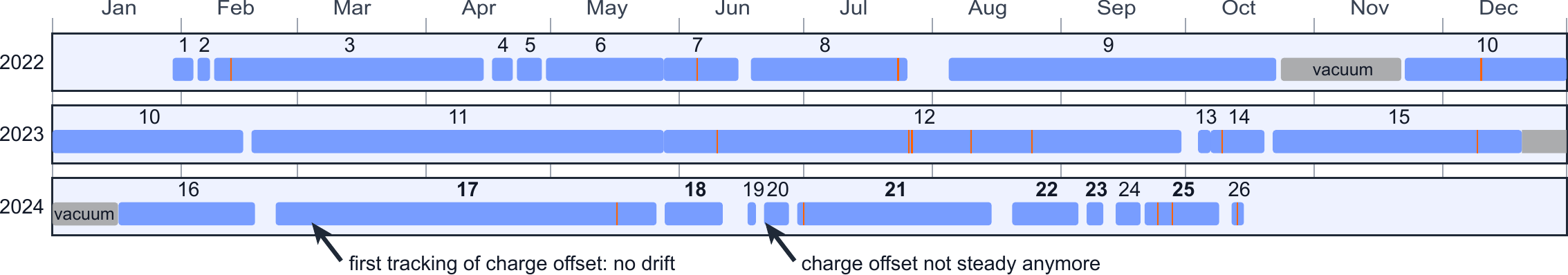}
    \caption{\textbf{Experiment history timeline.} Blue blocks represent run intervals labeled by a run number shown above each block. Bold text highlight the runs whose data are presented in this work. Orange lines represent thermal cycling events above 30~K. An absence of block represents an open dilution refrigerator at room temperature. 2025 was dedicated to surface analysis at room temperature.}
    \label{fig:history}
\end{figure*}
We measured the height difference between the pristine and etched regions with a Bruker Dektak profilometer. The expected step height was $\SI{200}{nm}$, corresponding to the nominal height of tantalum sputtered on the sapphire wafers. When performing four measurements close to the transmon cross, we reproducibly measured a step height of only $\SI{189(2)}{nm}$, as shown in \cref{fig:sup_dektak} (yellow curve). The tantalum film is thus mostly etched away in the nominal 17~s.

We also measured the step height on devices presented in \cref{fig:sup_xps} (orange and magenta in \cref{fig:sup_dektak}), and on another device made from the same wafer, hence the same sputtered tantalum layer, as for the device presented in the main text, but wet-etched separately (blue curve in \cref{fig:sup_dektak}). They all present a step height of $\SI{211(3)}{nm}$. This is a further clue that the wet etching of the device presented in the main text is incomplete. However, it seems very unlikely that there remains more than 20~nm of tantalum given the properties of the transmon and the results of the EDX analysis. We cannot exclude a scenario where the optical resist was lifted off during the etch and led to the partial etching of the tantalum pads themselves.

\section{History of the device}\label{sec:history}

The device was fabricated in 2021 for other experiments~\cite{Lledo2023, dassonneville_amplifying_2026}. Measurements probing the charge offset, obtained by tracking the $f_{34}$ transition frequency, began during \texttt{run 17}. In order to grasp the full history of the device, all the cooldowns and thermal cycles (TC) are represented in \cref{fig:history}.

On June 14, 2024 (before \texttt{run 19}), the device was inspected under an optical microscope, and the sample holder was mounted in a new can (see Supplementary \cref{sec:can}). \texttt{Run 19} was aborted due to a cabling inversion. After warmup at room temperature, the resistance between the charge line and the ground plane was measured and found to be above $20~\mathrm{M}\Omega$. Right before \texttt{run 25}, a wire bond was added between the charge line and the ground plane close to the edge of the chip. Right before \texttt{run 26}, a new wire bond was added to the charge line, close to the transmon cross-shape electrode. On October 8, 2024, the dilution refrigerator warmed up unexpectedly due to a cold leak from the dilution unit. The experiment stopped with \texttt{run 26} due to the leak. The surface analysis that followed damaged the Josephson junction of the transmon, which became an open circuit, thus preventing further low temperature measurements.

\begin{figure}
    \centering
    \includegraphics[width=1\columnwidth]{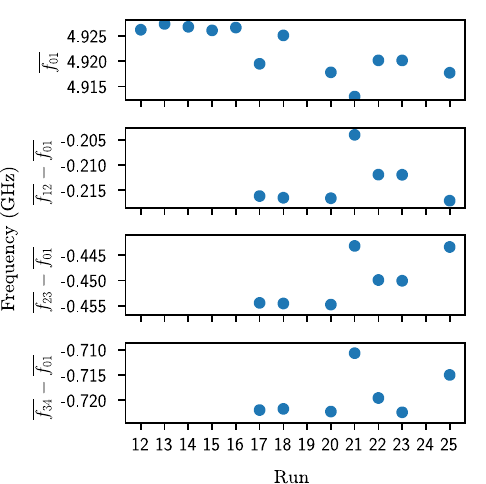}
    \caption{\textbf{Device spectrum of different runs.} Top panel: Measured mean transition frequency $\overline{f_{01}}$ as a function of the run. Bottom panels: Measured frequency difference between $\overline{f_{ii+1}}$ and $\overline{f_{01}}$.}
    \label{fig:freqchange}
\end{figure}

 The measured average transmon transition frequencies are shown in \cref{fig:freqchange} for \texttt{runs 12} to \texttt{25}. A prominent change of transition frequencies occured in \texttt{run 21} matching with the appearance of frequent noisy offset charge time intervals (Fig.~\ref{fig:4}a).

The qubit lifetime $T_1$ and coherence time $T_2$ were measured several times during the history of the sample. In Refs.~\cite{Lledo2023,dassonneville_amplifying_2026}, $T_1$ was measured in the range $20-\SI{27}{\micro s}$, and $T_2$ in the range $7.5-\SI{45}{\micro s}$. The qubit $T_1$ and $T_2$ of the runs used in this work are presented in \cref{tab:T1_T2}.

\begin{table}
    \centering
    \begin{tabular}{|c|c|c|}
         \hline
         Run&$T_1 (\SI{}{\micro s})$&$T_2 (\SI{}{\micro s})$\\
         \hline
         17&23-24&18-37\\
         18&21&15-30\\
         21&Not measured&43-45\\
         22&27&43-44\\
         23&26&22-23\\
         25&14-16&22-23\\
         \hline
    \end{tabular}
    \caption{Qubit lifetime $T_1$ and coherence time $T_2$ during the runs presented in this work.}
    \label{tab:T1_T2}
\end{table}

\begin{figure}
    \centering
    \includegraphics[width=1\columnwidth]{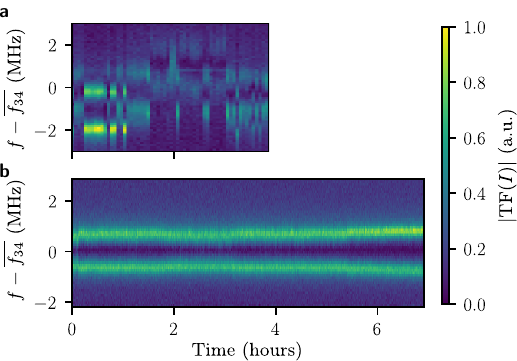}
    \caption{\textbf{Evolution of the charge offset during \texttt{run 22}.} \textbf{a}, Record of the frequency $f_{34}$  using the method of Fig.~\ref{fig:2} for 3.9 hours on August 24th, 2024. The repetition time is $\SI{139}{s}$. \textbf{b}, Record of the frequency $f_{34}$ for 6.9 hours on August 30th, 2024. The repetition time is $\SI{12}{s}$.}
    \label{fig:sup34_run22}
\end{figure}

At the beginning of \texttt{run 22}, an unusual time evolution of the frequency splitting was observed (\cref{fig:sup34_run22}a). The frequency splitting was mostly constant, but the mean value between the peaks was jumping several times per hour on seemingly discrete values. However, one week later (\cref{fig:sup34_run22}b), the frequency splitting was simply drifting as in the measurements presented \cref{fig:4}. We believe the initial behavior is due to the dynamics of a coupled TLS.

\section{Wire bond on the charge line}

\begin{figure}[h!]
    \centering
    \includegraphics[width=\columnwidth]{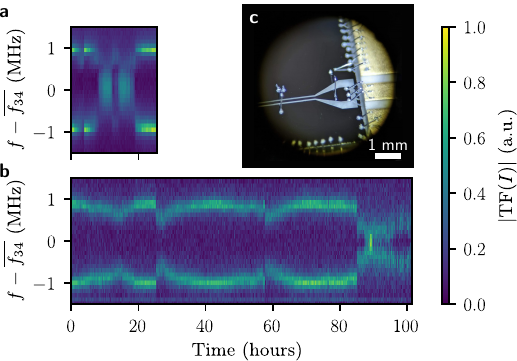}
    \caption{\textbf{Impact of a DC short from the charge line to the ground.} \textbf{a}, Record of the frequency $f_{34}$ during \texttt{run 24} for 1 full day using the technique of Fig.~\ref{fig:2}. The repetition time is $\SI{13}{s}$. \textbf{b}, Record of the frequency $f_{34}$ during \texttt{run 25} for 3.5 days. The repetition time is $\SI{13}{s}$.
    \textbf{c}, Optical image of the aluminum wire bond connecting the charge line and the ground plane in \texttt{run 25}.}
    \label{fig:sup_wire bond}
\end{figure}

In an attempt to recover the charge stability of the transmon once it began to drift, we connected the charge line to the ground with a wire bond, as shown in \cref{fig:sup_wire bond}c. With this wire bond in place, all metallic parts are connected by a low resistance path to the sample holder box, except for the cross-shaped island of the transmon. Interestingly, the few nH inductance of this aluminum wire bond is large enough to permit driving the qubit on resonance.

Comparing the drifts and jumps of $f_{34}$ with (\cref{fig:sup_wire bond}b) and without (\cref{fig:sup_wire bond}a) the wire bond, we do not observe a dramatic effect of the wire bond on the rate of charge offset jumps and drifts. We thus conclude that the origin of the exceptional charge stability in \texttt{runs 17} and {18} cannot be attributed to a DC short to ground of the charge line and readout resonator.

\section{Measurement setup}\label{sec:setup}
\begin{figure}[h!]
    \centering
    \includegraphics[width=\columnwidth]{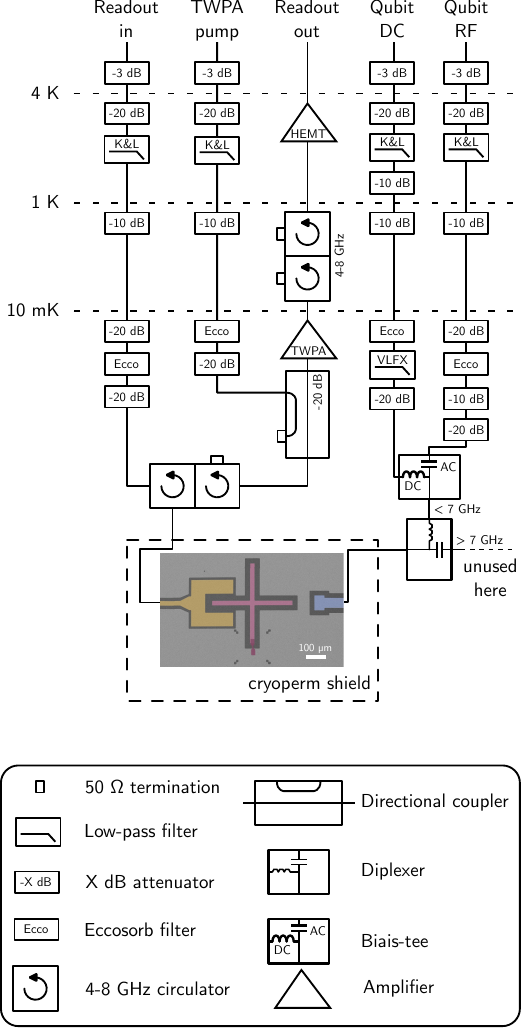}
    \caption{\textbf{Cabling diagram.} Schematic of the cabling inside the Bluefors LD250 dilution refrigerator used for the experiment with a base temperature at $\SI{10}{mK}$. The Josephson Traveling Wave Parametric Amplifier (TWPA) was provided by the Lincoln Lab.}
    \label{fig:schematic_fridge}
\end{figure}

The cryogenic microwave setup is shown in \cref{fig:schematic_fridge}. The qubit charge line is connected to a bias tee at the mixing chamber stage combining two different lines dedicated to low frequency (DC) and high frequency (RF) input signals. In fact, the charge line is also connected by a diplexer to another 50~$\Omega$ line not used in this experiment. The readout input pulses are attenuated up to a circulator before reaching the readout resonator. The reflected signal from the readout resonator goes through the same circulator and is then combined by a directional coupler with the pump tone of the Josephson Traveling Wave Parametric Amplifier (TWPA) \cite{Macklin2015} provided by the Lincoln Lab. The signal is then amplified by the TWPA, and a HEMT from Low Noise Factory. 

The qubit and readout pulses envelopes are generated at room temperature by an OPX from QuantumMachines acting as an Arbitrary Waveform Generator (AWG), with a sampling rate of $\SI{1}{GS/s}$. They are up-converted using IQ mixers by local oscillators generated by two APSIN12G Anapico. The TWPA pump is generated by an APSIN20G Anapico. The qubit DC voltage is applied by a BILT 2101 module. The readout output signal is down-converted at room temperature using image-reject mixers, amplified, and digitized by the OPX.

\label{sec:can}

\begin{figure}[h!]
    \centering
    \includegraphics[width=\columnwidth]{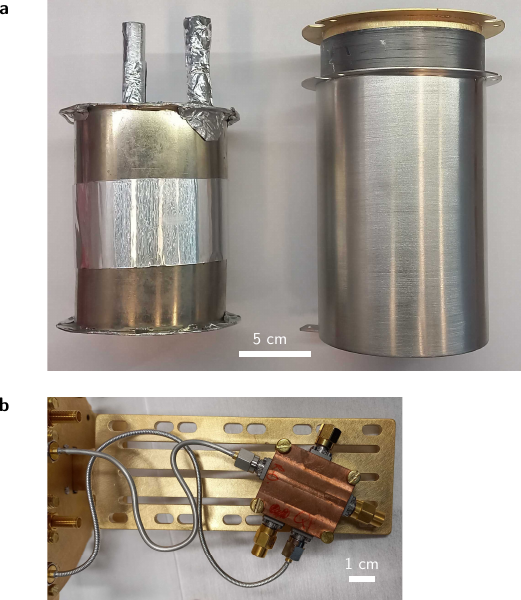}
    \caption{\textbf{Shielding of the device.} \textbf{a}, Can containing the device before (left) or after (right) June, 14th, 2024. The left can is made of a Cryoperm shield. The right can is made of a layer of gold-plated copper, a layer of lead and a layer of Cryoperm. \textbf{b}, Sample holder containing the sample.}
    \label{fig:sup_can}
\end{figure}

The shielding of the device is presented in \cref{fig:sup_can}. Until \texttt{run 19} on June 14, 2024, the sample holder was placed inside a can made of Cryoperm, and closed with aluminum tape. After that date, the sample holder was mounted in a can made of three layers: one layer of gold-plated copper, one layer of lead, and one layer of Cryoperm. The sample holder is shown in \cref{fig:sup_can}b.

\section{Investigation on the $2-3$ transition}

\begin{figure}[h!]
    \centering
    \includegraphics[width=\columnwidth]{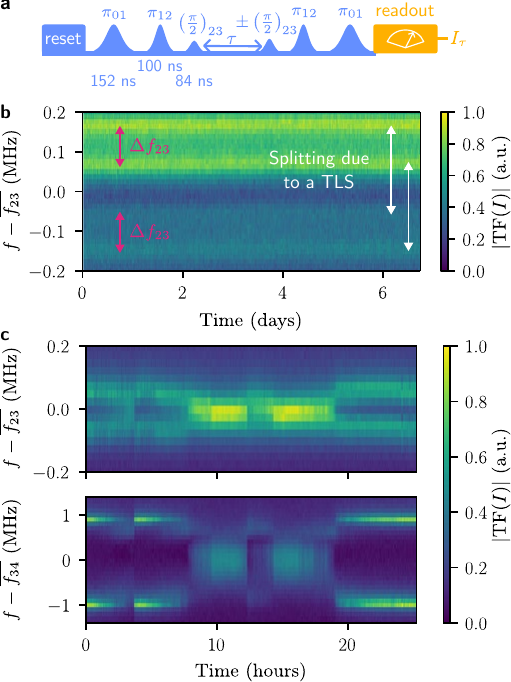}
    \caption{\textbf{Ramsey experiment on the $2-3$ transition.} \textbf{a}, Pulse sequence performing the measurement of Ramsey oscillations of the $2-3$ transition. Pink arrows: $f_{23}$ jumps by $\Delta f_{23}$ owing to charge parity switching. White arrows: $f_{23}$ jumps by a larger amount when a TLS changes state.\textbf{b}, Fourier transform of the signal $I_\tau$ output by the pulse sequence in a., as a function of time. The measurement was interleaved for 7 days during \texttt{run 17} with the one tracking $f_{34}$ shown in \cref{fig:3}. The repetition time is $\SI{58}{s}$. \textbf{c},  Continuous record of interleaved Ramsey experiments on the $2-3$ and $3-4$ transitions, measured from September 9 to 10, 2024 during \texttt{run 23}. The repetition time is $\SI{13}{s}$.}
    \label{fig:sup_23}
\end{figure}

The charge dispersion on the $2-3$ transition is measured to be $\Delta f_{23}=\SI{0.1}{MHz}$, it should thus also be possible to track charge drifts with this transition. We use the pulse sequence shown in \cref{fig:sup_23}a, which is similar to the one shown in \cref{fig:2}a.

During the week-long continuous measurement presented in \cref{fig:3}, interleaved Ramsey experiments were also performed on the $2-3$ transition (\cref{fig:sup_23}b). Four values of $f_{23}$ appear in this measurement. They are steady for the whole 7 days, consistent with the absence of charge-offset drift. The presence of four lines is attributed, as in \cref{fig:3}, to the two quasiparticle number parities and the two TLS states. In contrast to the $3-4$ transition, a change in the TLS state produces a larger shift of the $f_{23}$ frequency than a quasiparticle tunneling event. Note the similar impact on $f_{23}$ and $f_{34}$ of a change in TLS state (compare \cref{fig:3} and \cref{fig:sup_23}b).

Interleaved Ramsey experiments between the $2-3$ and $3-4$ experiments were also performed for a whole day during \texttt{run 23}, when the charge offset was drifting (\cref{fig:sup_23}c). We observe that the charge offset dynamics can be probed either on the $2-3$ or $3-4$ transitions. Since the frequency splitting on the $2-3$ transition is one order of magnitude smaller than that on the $3-4$ transition, the main text focuses on the latter. As expected, the jumps and drifts occur simultaneously for both transitions.

\section{Josephson harmonics and bound on the stray inductance}

The Hamiltonian describing the transmon and its coupled readout resonator
\begin{equation}
\begin{split}
\hat{H}_\mathrm{standard}=&\hbar\omega_r\hat{a}^\dagger\hat{a}+4 E_C(\hat{n}-n_g-n_\mathrm{qp}/2)^2\\
 & -E_J\cos\hat{\varphi}\\
 & +\hbar g (\hat{a}+\hat{a}^\dagger)\hat{n},\\
 \label{eq:fullH}
\end{split}
\end{equation}
fails to reproduce the observed higher transition frequencies $\overline{f_{ii+1}}$ and their charge dispersion $\Delta f_{ii+1}$ for \texttt{run 17}. Indeed, once the parameters of this Hamiltonian are set by matching the observed transition frequencies $f_{01}^\mathrm{even}$ and $f_{12}^\mathrm{even}\approx f_{01}^\mathrm{even}-E_C/h$ as well as the readout resonator frequencies when the qubit is in states $\ket{0}$ or $\ket{1}$, we find a large error of several MHz in the prediction of the transition frequencies. For instance, the predicted charge dispersion $\Delta f_{34}$ is half of the value we observe in the experiment. Even when we try to adapt the parameters to minimize the total error on the prediction of any transition frequencies we measured, discrepancies as large as a MHz remain. 

\begin{figure*}
    \centering
    \includegraphics[width=1\linewidth]{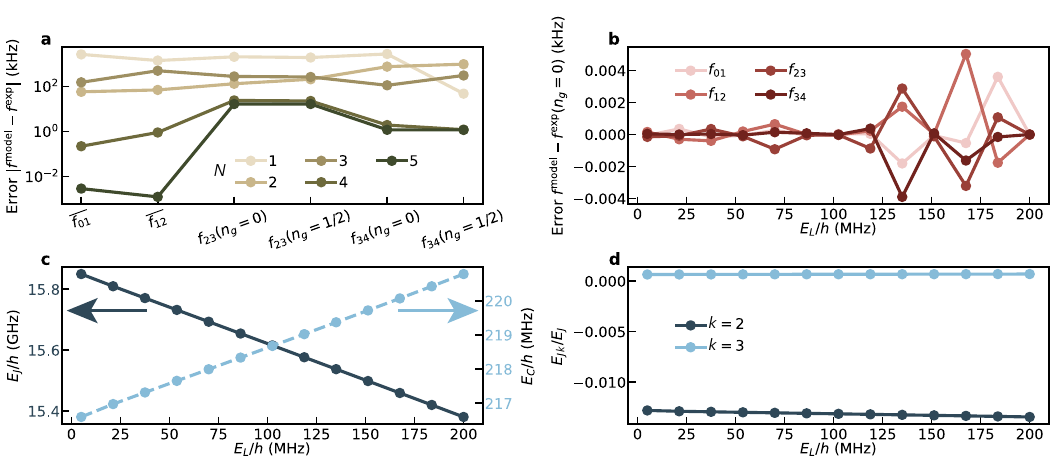}
    \caption{\textbf{Fitted parameters for extended transmon models.} \textbf{a} Deviation between measured transition frequencies (x-axis) during \texttt{run 17} and their best fit using the Hamiltonian \cref{eq:Hn} of a transmon with Josephson harmonics up to $1\leq N \leq 5$ (see colored legend). \textbf{b,} For each value of $E_L$, discrepancy between the transition frequencies $f_{ii+1}(n_g=0)$ and best fit of the spectrum $f_{ii+1}(n_g=0)$ of the Hamiltonian $\hat{H}_\mathrm{full}$ (\cref{eq:Hfull}). \textbf{c,d,} Fit parameters as a function of $E_L$.}
    \label{fig:harmonics}
\end{figure*}

\subsection{Josephson harmonics}
It has recently been discussed that the series inductance in the leads of the Josephson junction and possible highly transmitting channels in its oxide barrier lead to Josephson harmonics in the transmon Hamiltonian~\cite{willsch_observation_2024,kim_emergent_2025,wang_high-e_je_c_2025,shagalov_higher_2025,fechant_offset_2025,xia_exceeding_2025}, which then reads at order $N$ in these harmonics
\begin{equation}
\hat{H}_\mathrm{N}=4 E_C(\hat{n}-n_g-n_\mathrm{qp}/2)^2 -\sum_{k=1}^N E_{Jk}\cos(k\hat{\varphi}). 
 \label{eq:Hn}
\end{equation}
We note that $E_{J1}=E_J$ is the standard Josephson energy.

In \cref{fig:harmonics}, we show the smallest discrepancy between the measured transition frequencies in \texttt{run 17} and the ones predicted by this model for $1\leq N \leq 5$. For instance, the best fit to the observed transition frequencies we obtain with this model for $N=5$ gives the parameters $E_C/h= 216.5~\mathrm{MHz}$, $E_J/h= 15.6~\mathrm{GHz}$, $E_{J2}/h= -11.6~\mathrm{MHz}$, $E_{J3}/h= -122~\mathrm{MHz}$, $E_{J4}/h= 67.6~\mathrm{MHz}$, and $E_{J5}/h= -19.1~\mathrm{MHz}$.  The corresponding discrepancies between the measured and predicted transition frequencies are order of magnitude smaller than those with the standard model (darker versus lighter dots in \cref{fig:harmonics}a). To get a sense on the sensibility to the model parameters, note that the best fit for $N=3$ gives $E_C/h= 217.1~\mathrm{MHz}$, $E_J/h= 15.9~\mathrm{GHz}$, $E_{J2}/h= -227~\mathrm{MHz}$, $E_{J3}/h= -16.9~\mathrm{MHz}$. Adding the readout resonator to the model does not improve the fit.

Assuming this second harmonic entirely comes from a linear inductance $L_s$ in series with the Josephson junction, one expects~\cite{kim_emergent_2025}
\[L_s=\frac{E_{J2}\hbar^2}{E_J^2 e^2}.\]
Depending on the order $N$ in the harmonics, the inductance we predict from the best fit of the transition frequencies varies greatly. We find it from $30$ to $600~\mathrm{pH}$. It is compatible with the 80~pH that can be inferred from geometry and the known kinetic inductance of the 20~nm thick ($3$~pH per square) and 40~nm thick ($0.8$~pH per square) aluminum wires.

\subsection{Bound on the possible parallel inductance}
\label{sec:inductancebound}
\begin{figure*}
    \centering
    \includegraphics[width=1.0\linewidth]{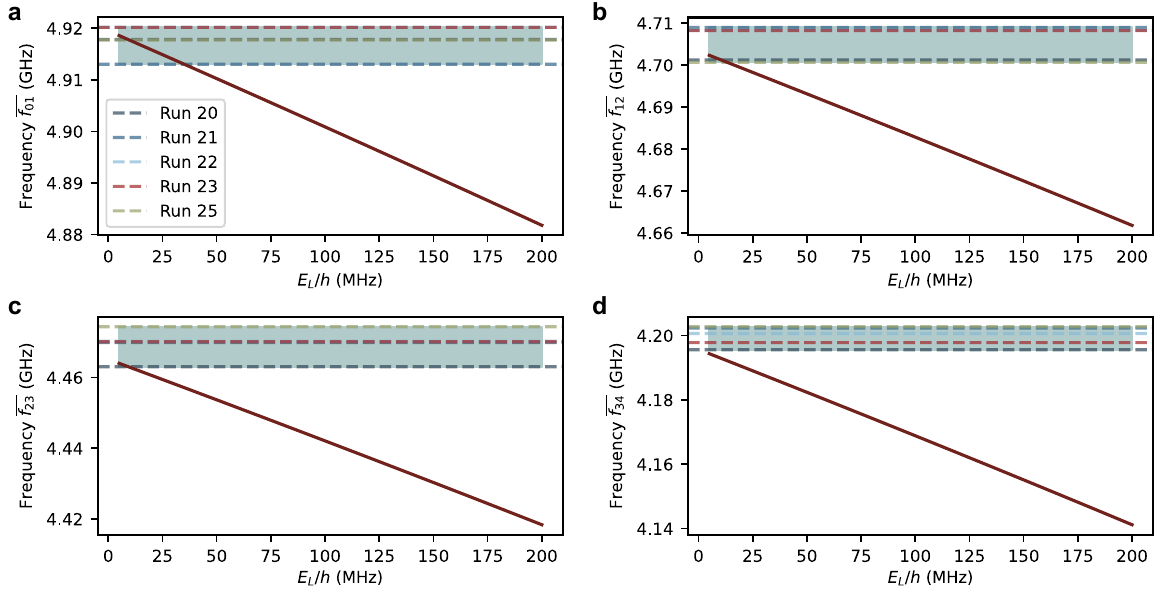}
    \caption{\textbf{Predicted shift in transition frequencies upon removal of the parallel inductance.}
    \textbf{a,} Solid dark red line: mean transition frequency $\overline{f_{01}}$ predicted by the model
    $\hat{H}_3$ (without parallel inductance) as a function of $E_L/h$, where the parameters $E_J$,
    $E_C$, $E_{J2}$, $E_{J3}$ are those obtained from fitting $\hat{H}_\mathrm{full}$ to the \texttt{run-17}
    spectrum at each value of $E_L$. This solid line therefore represents the frequency that would be
    measured after the inductance disappears, as a function of how large it was during \texttt{run 17}. Dashed
    lines: measured values of $\overline{f_{01}}$ in \texttt{runs 20-25}, after charge-offset stability was
    lost (see color legend). The shaded region spans the full spread of those measured values.
    \textbf{b,} Same for $\overline{f_{12}}$.
    \textbf{c,} Same for $\overline{f_{23}}$.
    \textbf{d,} Same for $\overline{f_{34}}$.}
    \label{fig:freqofEL}
\end{figure*}

\subsubsection{Bound coming from the circuit spectrum}

In this section, we compute the transmon spectrum in the case where an inductive element of inductance $L$ provides a path to ground in parallel with the junction. We will assume that the inductance can be modeled as a single branch, thus defining a loop threaded by a magnetic flux $\phi_\mathrm{ext}$. We use the following Hamiltonian to capture both the Josephson harmonics up to the third one and the extra inductive energy~\cite{Manucharyan2009,hassani_inductively_2023}:
\begin{equation}
\begin{split}
\hat{H}_\mathrm{full}=\hat{H}_3(n_g=0)+ \frac{1}{2} E_L (\hat{\phi} + \varphi_{\mathrm{ext}})^2,
 \label{eq:Hfull}
\end{split}
\end{equation}
where $\varphi_{\mathrm{ext}}=\phi_\mathrm{ext}/\varphi_0$ is the reduced flux through the loop, $\varphi_0=\frac{\hbar}{2e}$ and $E_L=\varphi_0^2/L$. Setting $\varphi_\mathrm{ext}=0$ for now, we identify, for each value of $E_L$, the parameters of the Hamiltonian \cref{eq:Hfull} that most closely reproduce all measured transition frequencies for $n_g=0$ and $n_\mathrm{qp}=0$. In \cref{fig:harmonics}b, one can see the minimal discrepancy we reach between the measured and computed transition frequencies as a function of $E_L$. The fit is excellent for any value of $E_L$ if all parameters are free, as expected since we do not use the information gathered at $n_\mathrm{qp}=1$. The corresponding values of $E_C$ and $E_J$ vary in opposite directions with $E_L$, and the harmonic terms $E_{J2}$ and $E_{J3}$ do not vary much as can be seen in \cref{fig:harmonics}c,d. We thus do not get a strong bound on $E_L$ from this analysis alone.

\subsubsection{Bound from the frequency shift upon disappearance of the inductance}

A constraint on the parallel inductance comes from comparing the transmon spectrum before and after charge-offset stability was lost. For each assumed value of $E_L$, the spectrum measured during \texttt{run 17} -- when the charge offset was pinned to zero -- uniquely determines the parameters
$E_J$, $E_C$, $E_{J2}$, and $E_{J3}$ of the full Hamiltonian $\hat{H}_\mathrm{full}$ (see \cref{fig:harmonics}c,d). If the inductance is removed from the circuit -- as may have occurred between \texttt{runs 18} and \texttt{20} -- the transition frequencies should shift to those predicted by $\hat{H}_3$ with those same parameters. The solid lines in \cref{fig:freqofEL} show these predicted frequencies as a function of $E_L$: the larger the inductance was during \texttt{run 17}, the larger the predicted frequency shift upon its disappearance.

The dashed lines and shaded region show the spread of transition frequencies actually measured in \texttt{runs 20} to \texttt{25}. These frequencies vary from run to run, but they remain confined to a narrow band. For $E_L/h \gtrsim 30~\mathrm{MHz}$ (equivalently $L \lesssim 20~\mathrm{\mu H}$), the predicted post-inductance frequencies fall well below this band, which is inconsistent with the observations. 

We therefore obtain a lower bound $L > 20~\mathrm{\mu H}$ on the parallel inductance present during \texttt{runs 17} and \texttt{18}.

\subsubsection{Bound coming from the flux dispersion}

Since the spectroscopy looks like that of a transmon, the inductive energy $E_L$ must be very small compared to $E_J$. In the regime $E_L\ll E_C\ll E_J$, the circuit transition frequencies for $n_g=n_\mathrm{qp}=0$ can be obtained in perturbation theory to second order~\cite{hassani_inductively_2023}. At the first order $N=1$ in the harmonics, second order in $\varphi_\mathrm{ext}$ and first orders in $E_C/E_J$ and $E_L/E_J$, we find
\begin{equation}
\begin{split}
f_{ii+1}(\varphi_\mathrm{ext}) = f_{ii+1}^\mathrm{\hat{H}_1} &+\frac{E_L }{h} \left( \frac{2E_C}{E_J} \right)^{\frac{1}{2}} \\
&\left(1-(i+1)\frac{E_L}{4E_J}-\varphi_\mathrm{ext}^2\frac{E_L}{4E_J}\right),
\end{split}
    \label{eq:transition}
\end{equation}
where we have used the fact that $f_{01}^\mathrm{\hat{H}_1}\approx \sqrt{8E_JE_C}$. Therefore, the transition frequency is expected to vary as
\begin{equation}
    \label{eq:transition01flux}
    f_{i i+1}(0)-f_{i i+1}(\varphi_\mathrm{ext})\approx\frac{\varphi_\mathrm{ext}^2}{8} \left( \frac{E_L}{E_J} \right)^2f_{01}^\mathrm{\hat{H}_1}.
\end{equation}

Assuming the flux takes arbitrary values from one cooldown to the next, the characteristic frequency change is found by setting $\varphi_\mathrm{ext}\approx 1$. We get
\begin{equation}
    f_{i i+1}(0)-f_{i i+1}(\varphi_\mathrm{ext})\approx\frac{1}{8} \left( \frac{E_L}{E_J} \right)^2f_{01}^\mathrm{\hat{H}_1}.
\end{equation}
Thus, the frequency change of the order of $\SI{5}{MHz}$ from \texttt{run 17} to \texttt{run 18} is compatible with 
\begin{equation}
    \label{eq:ELmax}
    \frac{E_L}h<\frac{E_J}h\sqrt{\frac{8\cdot 5~\mathrm{MHz}}{f_{01}^\mathrm{\hat{H}_1}}}\approx 1.4~\mathrm{GHz}\Leftrightarrow L>0.47~\mathrm{\mu H}.
\end{equation}
We conclude that flux dispersion does not offer the strongest bound on $E_L$.



\begin{thebibliography}{59}
\ifx \bisbn   \undefined \def \bisbn  #1{ISBN #1}\fi
\ifx \binits  \undefined \def \binits#1{#1}\fi
\ifx \bauthor  \undefined \def \bauthor#1{#1}\fi
\ifx \batitle  \undefined \def \batitle#1{#1}\fi
\ifx \bjtitle  \undefined \def \bjtitle#1{#1}\fi
\ifx \bvolume  \undefined \def \bvolume#1{\textbf{#1}}\fi
\ifx \byear  \undefined \def \byear#1{#1}\fi
\ifx \bissue  \undefined \def \bissue#1{#1}\fi
\ifx \bfpage  \undefined \def \bfpage#1{#1}\fi
\ifx \blpage  \undefined \def \blpage #1{#1}\fi
\ifx \burl  \undefined \def \burl#1{\textsf{#1}}\fi
\ifx \doiurl  \undefined \def \doiurl#1{\url{https://doi.org/#1}}\fi
\ifx \betal  \undefined \def \betal{\textit{et al.}}\fi
\ifx \binstitute  \undefined \def \binstitute#1{#1}\fi
\ifx \binstitutionaled  \undefined \def \binstitutionaled#1{#1}\fi
\ifx \bctitle  \undefined \def \bctitle#1{#1}\fi
\ifx \beditor  \undefined \def \beditor#1{#1}\fi
\ifx \bpublisher  \undefined \def \bpublisher#1{#1}\fi
\ifx \bbtitle  \undefined \def \bbtitle#1{#1}\fi
\ifx \bedition  \undefined \def \bedition#1{#1}\fi
\ifx \bseriesno  \undefined \def \bseriesno#1{#1}\fi
\ifx \blocation  \undefined \def \blocation#1{#1}\fi
\ifx \bsertitle  \undefined \def \bsertitle#1{#1}\fi
\ifx \bsnm \undefined \def \bsnm#1{#1}\fi
\ifx \bsuffix \undefined \def \bsuffix#1{#1}\fi
\ifx \bparticle \undefined \def \bparticle#1{#1}\fi
\ifx \barticle \undefined \def \barticle#1{#1}\fi
\bibcommenthead
\ifx \bconfdate \undefined \def \bconfdate #1{#1}\fi
\ifx \botherref \undefined \def \botherref #1{#1}\fi
\ifx \url \undefined \def \url#1{\textsf{#1}}\fi
\ifx \bchapter \undefined \def \bchapter#1{#1}\fi
\ifx \bbook \undefined \def \bbook#1{#1}\fi
\ifx \bcomment \undefined \def \bcomment#1{#1}\fi
\ifx \oauthor \undefined \def \oauthor#1{#1}\fi
\ifx \citeauthoryear \undefined \def \citeauthoryear#1{#1}\fi
\ifx \endbibitem  \undefined \def \endbibitem {}\fi
\ifx \bconflocation  \undefined \def \bconflocation#1{#1}\fi
\ifx \arxivurl  \undefined \def \arxivurl#1{\textsf{#1}}\fi
\csname PreBibitemsHook\endcsname

\bibitem{dutta_low-frequency_1981}
\begin{barticle}
\bauthor{\bsnm{Dutta}, \binits{P.}},
\bauthor{\bsnm{Horn}, \binits{P.M.}}:
\batitle{Low-frequency fluctuations in solids: 1/f noise}.
\bjtitle{Reviews of Modern Physics}
\bvolume{53}(\bissue{3}),
\bfpage{497}--\blpage{516}
(\byear{1981}).
\doiurl{10.1103/RevModPhys.53.497}
\end{barticle}
\endbibitem

\bibitem{kogan_electronic_1996}
\begin{bbook}
\bauthor{\bsnm{Kogan}, \binits{S.}}:
\bbtitle{Electronic {Noise} and {Fluctuations} in {Solids}}.
\bpublisher{Cambridge University Press},
\blocation{Cambridge}
(\byear{1996})
\end{bbook}
\endbibitem

\bibitem{weissman_frac1f_1988}
\begin{barticle}
\bauthor{\bsnm{Weissman}, \binits{M.B.}}:
\batitle{1/f noise and other slow, nonexponential kinetics in condensed matter}.
\bjtitle{Reviews of Modern Physics}
\bvolume{60}(\bissue{2}),
\bfpage{537}--\blpage{571}
(\byear{1988}).
\doiurl{10.1103/RevModPhys.60.537}
\end{barticle}
\endbibitem

\bibitem{koelle_high-transition-temperature_1999}
\begin{barticle}
\bauthor{\bsnm{Koelle}, \binits{D.}},
\bauthor{\bsnm{Kleiner}, \binits{R.}},
\bauthor{\bsnm{Ludwig}, \binits{F.}},
\bauthor{\bsnm{Dantsker}, \binits{E.}},
\bauthor{\bsnm{Clarke}, \binits{J.}}:
\batitle{High-transition-temperature superconducting quantum interference devices}.
\bjtitle{Reviews of Modern Physics}
\bvolume{71}(\bissue{3}),
\bfpage{631}--\blpage{686}
(\byear{1999}).
\doiurl{10.1103/RevModPhys.71.631}
\end{barticle}
\endbibitem

\bibitem{grasser_noise_2020}
\begin{bbook}
\beditor{\bsnm{Grasser}, \binits{T.}} (ed.):
\bbtitle{Noise in {Nanoscale} {Semiconductor} {Devices}}.
\bpublisher{Springer},
\blocation{Cham}
(\byear{2020}).
\doiurl{10.1007/978-3-030-37500-3}
\end{bbook}
\endbibitem

\bibitem{wilen_correlated_2021}
\begin{barticle}
\bauthor{\bsnm{Wilen}, \binits{C.D.}},
\bauthor{\bsnm{Abdullah}, \binits{S.}},
\bauthor{\bsnm{Kurinsky}, \binits{N.A.}},
\bauthor{\bsnm{Stanford}, \binits{C.}},
\bauthor{\bsnm{Cardani}, \binits{L.}},
\bauthor{\bsnm{D’Imperio}, \binits{G.}},
\bauthor{\bsnm{Tomei}, \binits{C.}},
\bauthor{\bsnm{Faoro}, \binits{L.}},
\bauthor{\bsnm{Ioffe}, \binits{L.B.}},
\bauthor{\bsnm{Liu}, \binits{C.H.}},
\bauthor{\bsnm{Opremcak}, \binits{A.}},
\bauthor{\bsnm{Christensen}, \binits{B.G.}},
\bauthor{\bsnm{DuBois}, \binits{J.L.}},
\bauthor{\bsnm{McDermott}, \binits{R.}}:
\batitle{Correlated charge noise and relaxation errors in superconducting qubits}.
\bjtitle{Nature}
\bvolume{594}(\bissue{7863}),
\bfpage{369}--\blpage{373}
(\byear{2021}).
\doiurl{10.1038/s41586-021-03557-5}
\end{barticle}
\endbibitem

\bibitem{thorbeck_two-level-system_2023}
\begin{barticle}
\bauthor{\bsnm{Thorbeck}, \binits{T.}},
\bauthor{\bsnm{Eddins}, \binits{A.}},
\bauthor{\bsnm{Lauer}, \binits{I.}},
\bauthor{\bsnm{McClure}, \binits{D.T.}},
\bauthor{\bsnm{Carroll}, \binits{M.}}:
\batitle{Two-{Level}-{System} {Dynamics} in a {Superconducting} {Qubit} {Due} to {Background} {Ionizing} {Radiation}}.
\bjtitle{PRX Quantum}
\bvolume{4}(\bissue{2}),
\bfpage{020356}
(\byear{2023}).
\doiurl{10.1103/PRXQuantum.4.020356}
\end{barticle}
\endbibitem

\bibitem{falci_1f_2024}
\begin{botherref}
\oauthor{\bsnm{Falci}, \binits{G.}},
\oauthor{\bsnm{Hakonen}, \binits{P.J.}},
\oauthor{\bsnm{Paladino}, \binits{E.}}:
1/f noise in quantum nanoscience.
In: Chakraborty, T. (ed.)
Encyclopedia of {Condensed} {Matter} {Physics} ({Second} {Edition}),
Oxford,
pp. 1003--1017
(2024).
\doiurl{10.1016/B978-0-323-90800-9.00250-X}
\end{botherref}
\endbibitem

\bibitem{larson_quasiparticle_2025}
\begin{barticle}
\bauthor{\bsnm{Larson}, \binits{C.P.}},
\bauthor{\bsnm{Yelton}, \binits{E.}},
\bauthor{\bsnm{Dodge}, \binits{K.}},
\bauthor{\bsnm{Okubo}, \binits{K.}},
\bauthor{\bsnm{Batarekh}, \binits{J.}},
\bauthor{\bsnm{Iaia}, \binits{V.}},
\bauthor{\bsnm{Kurinsky}, \binits{N.A.}},
\bauthor{\bsnm{Plourde}, \binits{B.L.T.}}:
\batitle{Quasiparticle poisoning of superconducting qubits with active gamma irradiation}.
\bjtitle{PRX Quantum}
\bvolume{6},
\bfpage{030339}
(\byear{2025}).
\doiurl{10.1103/2lyd-8swv}
\end{barticle}
\endbibitem

\bibitem{Nakamura1999}
\begin{barticle}
\bauthor{\bsnm{Nakamura}, \binits{Y.}},
\bauthor{\bsnm{Pashkin}, \binits{Y.A.}},
\bauthor{\bsnm{Tsai}, \binits{J.S.}}:
\batitle{Coherent control of macroscopic quantum states in a single-{Cooper}-pair box}.
\bjtitle{Nature}
\bvolume{398}(\bissue{6730}),
\bfpage{786}--\blpage{788}
(\byear{1999}).
\doiurl{10.1038/19718}
\end{barticle}
\endbibitem

\bibitem{Vion2002}
\begin{barticle}
\bauthor{\bsnm{Vion}, \binits{D.}},
\bauthor{\bsnm{Aassime}, \binits{a.}},
\bauthor{\bsnm{Cottet}, \binits{a.}},
\bauthor{\bsnm{Joyez}, \binits{P.}},
\bauthor{\bsnm{Pothier}, \binits{H.}},
\bauthor{\bsnm{Urbina}, \binits{C.}},
\bauthor{\bsnm{Esteve}, \binits{D.}},
\bauthor{\bsnm{Devoret}, \binits{M.H.}}:
\batitle{Manipulating the quantum state of an electrical circuit.}
\bjtitle{Science (New York, N.Y.)}
\bvolume{296}(\bissue{5569}),
\bfpage{886}--\blpage{9}
(\byear{2002}).
\doiurl{10.1126/science.1069372}
\end{barticle}
\endbibitem

\bibitem{bladh_single_2005}
\begin{barticle}
\bauthor{\bsnm{Bladh}, \binits{K.}},
\bauthor{\bsnm{Duty}, \binits{T.}},
\bauthor{\bsnm{Gunnarsson}, \binits{D.}},
\bauthor{\bsnm{Delsing}, \binits{P.}}:
\batitle{The single {Cooper}-pair box as a charge qubit}.
\bjtitle{New Journal of Physics}
\bvolume{7}(\bissue{1}),
\bfpage{180}
(\byear{2005}).
\doiurl{10.1088/1367-2630/7/1/180}
\end{barticle}
\endbibitem

\bibitem{metcalfe_measuring_2007}
\begin{barticle}
\bauthor{\bsnm{Metcalfe}, \binits{M.}},
\bauthor{\bsnm{Boaknin}, \binits{E.}},
\bauthor{\bsnm{Manucharyan}, \binits{V.}},
\bauthor{\bsnm{Vijay}, \binits{R.}},
\bauthor{\bsnm{Siddiqi}, \binits{I.}},
\bauthor{\bsnm{Rigetti}, \binits{C.}},
\bauthor{\bsnm{Frunzio}, \binits{L.}},
\bauthor{\bsnm{Schoelkopf}, \binits{R.J.}},
\bauthor{\bsnm{Devoret}, \binits{M.H.}}:
\batitle{Measuring the decoherence of a quantronium qubit with the cavity bifurcation amplifier}.
\bjtitle{Physical Review B}
\bvolume{76}(\bissue{17}),
\bfpage{174516}
(\byear{2007}).
\doiurl{10.1103/PhysRevB.76.174516}
\end{barticle}
\endbibitem

\bibitem{Koch2007}
\begin{barticle}
\bauthor{\bsnm{Koch}, \binits{J.}},
\bauthor{\bsnm{Yu}, \binits{T.M.}},
\bauthor{\bsnm{Gambetta}, \binits{J.}},
\bauthor{\bsnm{Houck}, \binits{A.A.}},
\bauthor{\bsnm{Schuster}, \binits{D.I.}},
\bauthor{\bsnm{Majer}, \binits{J.}},
\bauthor{\bsnm{Blais}, \binits{A.}},
\bauthor{\bsnm{Devoret}, \binits{M.H.}},
\bauthor{\bsnm{Girvin}, \binits{S.M.}},
\bauthor{\bsnm{Schoelkopf}, \binits{R.J.}}:
\batitle{Charge-insensitive qubit design derived from the {Cooper} pair box}.
\bjtitle{Phys. Rev. A}
\bvolume{76},
\bfpage{042319}
(\byear{2007}).
\doiurl{10.1103/PhysRevA.76.042319}
\end{barticle}
\endbibitem

\bibitem{Manucharyan2009}
\begin{barticle}
\bauthor{\bsnm{Manucharyan}, \binits{V.E.}},
\bauthor{\bsnm{Koch}, \binits{J.}},
\bauthor{\bsnm{Glazman}, \binits{L.I.}},
\bauthor{\bsnm{Devoret}, \binits{M.H.}}:
\batitle{Fluxonium: {Single} {Cooper}-{Pair} {Circuit} {Free} of {Charge} {Offsets}}.
\bjtitle{Science}
\bvolume{326}(\bissue{5949}),
\bfpage{113}--\blpage{116}
(\byear{2009}).
\doiurl{10.1126/science.1175552}
\end{barticle}
\endbibitem

\bibitem{hassani_inductively_2023}
\begin{barticle}
\bauthor{\bsnm{Hassani}, \binits{F.}},
\bauthor{\bsnm{Peruzzo}, \binits{M.}},
\bauthor{\bsnm{Kapoor}, \binits{L.N.}},
\bauthor{\bsnm{Trioni}, \binits{A.}},
\bauthor{\bsnm{Zemlicka}, \binits{M.}},
\bauthor{\bsnm{Fink}, \binits{J.M.}}:
\batitle{Inductively shunted transmons exhibit noise insensitive plasmon states and a fluxon decay exceeding 3 hours}.
\bjtitle{Nature Communications}
\bvolume{14}(\bissue{1}),
\bfpage{3968}
(\byear{2023}).
\doiurl{10.1038/s41467-023-39656-2}
\end{barticle}
\endbibitem

\bibitem{khezri_measurement-induced_2023}
\begin{barticle}
\bauthor{\bsnm{Khezri}, \binits{M.}},
\bauthor{\bsnm{Opremcak}, \binits{A.}},
\bauthor{\bsnm{Chen}, \binits{Z.}},
\bauthor{\bsnm{Miao}, \binits{K.C.}},
\bauthor{\bsnm{McEwen}, \binits{M.}},
\bauthor{\bsnm{Bengtsson}, \binits{A.}},
\bauthor{\bsnm{White}, \binits{T.}},
\bauthor{\bsnm{Naaman}, \binits{O.}},
\bauthor{\bsnm{Sank}, \binits{D.}},
\bauthor{\bsnm{Korotkov}, \binits{A.N.}},
\bauthor{\bsnm{Chen}, \binits{Y.}},
\bauthor{\bsnm{Smelyanskiy}, \binits{V.}}:
\batitle{Measurement-induced state transitions in a superconducting qubit: {Within} the rotating-wave approximation}.
\bjtitle{Physical Review Applied}
\bvolume{20}(\bissue{5}),
\bfpage{054008}
(\byear{2023}).
\doiurl{10.1103/PhysRevApplied.20.054008}
\end{barticle}
\endbibitem

\bibitem{cohen_reminiscence_2023}
\begin{barticle}
\bauthor{\bsnm{Cohen}, \binits{J.}},
\bauthor{\bsnm{Petrescu}, \binits{A.}},
\bauthor{\bsnm{Shillito}, \binits{R.}},
\bauthor{\bsnm{Blais}, \binits{A.}}:
\batitle{Reminiscence of {Classical} {Chaos} in {Driven} {Transmons}}.
\bjtitle{PRX Quantum}
\bvolume{4}(\bissue{2}),
\bfpage{020312}
(\byear{2023}).
\doiurl{10.1103/PRXQuantum.4.020312}
\end{barticle}
\endbibitem

\bibitem{dumas_measurement-induced_2024}
\begin{barticle}
\bauthor{\bsnm{Dumas}, \binits{M.F.}},
\bauthor{\bsnm{Groleau-Paré}, \binits{B.}},
\bauthor{\bsnm{McDonald}, \binits{A.}},
\bauthor{\bsnm{Muñoz-Arias}, \binits{M.H.}},
\bauthor{\bsnm{Lledó}, \binits{C.}},
\bauthor{\bsnm{D’Anjou}, \binits{B.}},
\bauthor{\bsnm{Blais}, \binits{A.}}:
\batitle{Measurement-{Induced} {Transmon} {Ionization}}.
\bjtitle{Physical Review X}
\bvolume{14}(\bissue{4}),
\bfpage{041023}
(\byear{2024}).
\doiurl{10.1103/PhysRevX.14.041023}
\end{barticle}
\endbibitem

\bibitem{fechant_offset_2025}
\begin{barticle}
\bauthor{\bsnm{Féchant}, \binits{M.}},
\bauthor{\bsnm{Dumas}, \binits{M.F.}},
\bauthor{\bsnm{Bénâtre}, \binits{D.}},
\bauthor{\bsnm{Gosling}, \binits{N.}},
\bauthor{\bsnm{Lenhard}, \binits{P.}},
\bauthor{\bsnm{Spiecker}, \binits{M.}},
\bauthor{\bsnm{Geisert}, \binits{S.}},
\bauthor{\bsnm{Ihssen}, \binits{S.}},
\bauthor{\bsnm{Wernsdorfer}, \binits{W.}},
\bauthor{\bsnm{D’Anjou}, \binits{B.}},
\bauthor{\bsnm{Blais}, \binits{A.}},
\bauthor{\bsnm{Pop}, \binits{I.M.}}:
\batitle{Offset {Charge} {Dependence} of {Measurement}-{Induced} {Transitions} in {Transmons}}.
\bjtitle{Physical Review Letters}
\bvolume{135}(\bissue{18}),
\bfpage{180603}
(\byear{2025}).
\doiurl{10.1103/yljv-b4kj}
\end{barticle}
\endbibitem

\bibitem{wang_probing_2025}
\begin{botherref}
\oauthor{\bsnm{Wang}, \binits{Z.}},
\oauthor{\bsnm{D'Anjou}, \binits{B.}},
\oauthor{\bsnm{Gigon}, \binits{P.}},
\oauthor{\bsnm{Blais}, \binits{A.}},
\oauthor{\bsnm{Blok}, \binits{M.S.}}:
Probing excited-state dynamics of transmon ionization
(2025)
{\href{https://arxiv.org/abs/2505.00639}{{arXiv:2505.00639}}}
{[quant-ph]}
\end{botherref}
\endbibitem

\bibitem{groszkowski_coherence_2018}
\begin{barticle}
\bauthor{\bsnm{Groszkowski}, \binits{P.}},
\bauthor{\bsnm{Paolo}, \binits{A.D.}},
\bauthor{\bsnm{Grimsmo}, \binits{A.L.}},
\bauthor{\bsnm{Blais}, \binits{A.}},
\bauthor{\bsnm{Schuster}, \binits{D.I.}},
\bauthor{\bsnm{Houck}, \binits{A.A.}},
\bauthor{\bsnm{Koch}, \binits{J.}}:
\batitle{Coherence properties of the 0-$\pi$ qubit}.
\bjtitle{New Journal of Physics}
\bvolume{20}(\bissue{4}),
\bfpage{043053}
(\byear{2018}).
\doiurl{10.1088/1367-2630/aab7cd}
\end{barticle}
\endbibitem

\bibitem{Riste2013}
\begin{barticle}
\bauthor{\bsnm{Rist{\`e}}, \binits{D.}},
\bauthor{\bsnm{Bultink}, \binits{C.}},
\bauthor{\bsnm{Tiggelman}, \binits{M.J.}},
\bauthor{\bsnm{Schouten}, \binits{R.N.}},
\bauthor{\bsnm{Lehnert}, \binits{K.W.}},
\bauthor{\bsnm{DiCarlo}, \binits{L.}}:
\batitle{Millisecond charge-parity fluctuations and induced decoherence in a superconducting transmon qubit}.
\bjtitle{Nature communications}
\bvolume{4}(\bissue{1}),
\bfpage{1913}
(\byear{2013}).
\doiurl{10.1038/ncomms2936}
\end{barticle}
\endbibitem

\bibitem{serniak_hot_2018}
\begin{botherref}
\oauthor{\bsnm{Serniak}, \binits{K.}},
\oauthor{\bsnm{Hays}, \binits{M.}},
\oauthor{\bsnm{De~Lange}, \binits{G.}},
\oauthor{\bsnm{Diamond}, \binits{S.}},
\oauthor{\bsnm{Shankar}, \binits{S.}},
\oauthor{\bsnm{Burkhart}, \binits{L.D.}},
\oauthor{\bsnm{Frunzio}, \binits{L.}},
\oauthor{\bsnm{Houzet}, \binits{M.}},
\oauthor{\bsnm{Devoret}, \binits{M.H.}}:
Hot {Nonequilibrium} {Quasiparticles} in {Transmon} {Qubits}.
Physical Review Letters
\textbf{121}(15)
(2018).
\doiurl{10.1103/PhysRevLett.121.157701}
\end{botherref}
\endbibitem

\bibitem{christensen_anomalous_2019}
\begin{botherref}
\oauthor{\bsnm{Christensen}, \binits{B.G.}},
\oauthor{\bsnm{Wilen}, \binits{C.D.}},
\oauthor{\bsnm{Opremcak}, \binits{A.}},
\oauthor{\bsnm{Nelson}, \binits{J.}},
\oauthor{\bsnm{Schlenker}, \binits{F.}},
\oauthor{\bsnm{Zimonick}, \binits{C.H.}},
\oauthor{\bsnm{Faoro}, \binits{L.}},
\oauthor{\bsnm{Ioffe}, \binits{L.B.}},
\oauthor{\bsnm{Rosen}, \binits{Y.J.}},
\oauthor{\bsnm{Dubois}, \binits{J.L.}},
\oauthor{\bsnm{Plourde}, \binits{B.L.T.}},
\oauthor{\bsnm{Mcdermott}, \binits{R.}}:
Anomalous charge noise in superconducting qubits.
Physical Review B
\textbf{100}(14)
(2019).
\doiurl{10.1103/PhysRevB.100.140503}
\end{botherref}
\endbibitem

\bibitem{serniak_direct_2019}
\begin{botherref}
\oauthor{\bsnm{Serniak}, \binits{K.}},
\oauthor{\bsnm{Diamond}, \binits{S.}},
\oauthor{\bsnm{Hays}, \binits{M.}},
\oauthor{\bsnm{Fatemi}, \binits{V.}},
\oauthor{\bsnm{Shankar}, \binits{S.}},
\oauthor{\bsnm{Frunzio}, \binits{L.}},
\oauthor{\bsnm{Schoelkopf}, \binits{R.J.}},
\oauthor{\bsnm{Devoret}, \binits{M.H.}}:
Direct {Dispersive} {Monitoring} of {Charge} {Parity} in {Offset}-{Charge}-{Sensitive} {Transmons}.
Physical Review Applied
\textbf{12}(1)
(2019).
\doiurl{10.1103/PhysRevApplied.12.014052}
\end{botherref}
\endbibitem

\bibitem{Tomonaga2021}
\begin{barticle}
\bauthor{\bsnm{Tomonaga}, \binits{A.}},
\bauthor{\bsnm{Mukai}, \binits{H.}},
\bauthor{\bsnm{Yoshihara}, \binits{F.}},
\bauthor{\bsnm{Tsai}, \binits{J.S.}}:
\batitle{Quasiparticle tunneling and $1/f$ charge noise in ultrastrongly coupled superconducting qubit and resonator}.
\bjtitle{Phys. Rev. B}
\bvolume{104},
\bfpage{224509}
(\byear{2021}).
\doiurl{10.1103/PhysRevB.104.224509}
\end{barticle}
\endbibitem

\bibitem{iaia_phonon_2022}
\begin{botherref}
\oauthor{\bsnm{Iaia}, \binits{V.}},
\oauthor{\bsnm{Ku}, \binits{J.}},
\oauthor{\bsnm{Ballard}, \binits{A.}},
\oauthor{\bsnm{Larson}, \binits{C.P.}},
\oauthor{\bsnm{Yelton}, \binits{E.}},
\oauthor{\bsnm{Liu}, \binits{C.H.}},
\oauthor{\bsnm{Patel}, \binits{S.}},
\oauthor{\bsnm{McDermott}, \binits{R.}},
\oauthor{\bsnm{Plourde}, \binits{B.L.T.}}:
Phonon downconversion to suppress correlated errors in superconducting qubits.
Nature Communications
\textbf{13}(1)
(2022).
\doiurl{10.1038/s41467-022-33997-0}
\end{botherref}
\endbibitem

\bibitem{pan_engineering_2022}
\begin{botherref}
\oauthor{\bsnm{Pan}, \binits{X.}},
\oauthor{\bsnm{Zhou}, \binits{Y.}},
\oauthor{\bsnm{Yuan}, \binits{H.}},
\oauthor{\bsnm{Nie}, \binits{L.}},
\oauthor{\bsnm{Wei}, \binits{W.}},
\oauthor{\bsnm{Zhang}, \binits{L.}},
\oauthor{\bsnm{Li}, \binits{J.}},
\oauthor{\bsnm{Liu}, \binits{S.}},
\oauthor{\bsnm{Jiang}, \binits{Z.H.}},
\oauthor{\bsnm{Catelani}, \binits{G.}},
\oauthor{\bsnm{Hu}, \binits{L.}},
\oauthor{\bsnm{Yan}, \binits{F.}},
\oauthor{\bsnm{Yu}, \binits{D.}}:
Engineering superconducting qubits to reduce quasiparticles and charge noise.
Nature Communications
\textbf{13}(1)
(2022).
\doiurl{10.1038/s41467-022-34727-2}
\end{botherref}
\endbibitem

\bibitem{tennant_low-frequency_2022}
\begin{barticle}
\bauthor{\bsnm{Tennant}, \binits{D.M.}},
\bauthor{\bsnm{Martinez}, \binits{L.A.}},
\bauthor{\bsnm{Beck}, \binits{K.M.}},
\bauthor{\bsnm{O'Kelley}, \binits{S.R.}},
\bauthor{\bsnm{Wilen}, \binits{C.D.}},
\bauthor{\bsnm{McDermott}, \binits{R.}},
\bauthor{\bsnm{DuBois}, \binits{J.L.}},
\bauthor{\bsnm{Rosen}, \binits{Y.J.}}:
\batitle{Low-{Frequency} {Correlated} {Charge}-{Noise} {Measurements} {Across} {Multiple} {Energy} {Transitions} in a {Tantalum} {Transmon}}.
\bjtitle{PRX Quantum}
\bvolume{3}(\bissue{3}),
\bfpage{030307}
(\byear{2022}).
\doiurl{10.1103/PRXQuantum.3.030307}
\end{barticle}
\endbibitem

\bibitem{wills_spatial_2022}
\begin{botherref}
\oauthor{\bsnm{Wills}, \binits{J.}},
\oauthor{\bsnm{Campanaro}, \binits{G.}},
\oauthor{\bsnm{Cao}, \binits{S.}},
\oauthor{\bsnm{Fasciati}, \binits{S.D.}},
\oauthor{\bsnm{Leek}, \binits{P.J.}},
\oauthor{\bsnm{Vlastakis}, \binits{B.}}:
Spatial {Charge} {Sensitivity} in a {Multimode} {Superconducting} {Qubit}.
Physical Review Applied
\textbf{17}(2)
(2022).
\doiurl{10.1103/PhysRevApplied.17.024058}
\end{botherref}
\endbibitem

\bibitem{martinez_noise-specific_2023}
\begin{botherref}
\oauthor{\bsnm{Martinez}, \binits{L.A.}},
\oauthor{\bsnm{Peng}, \binits{Z.}},
\oauthor{\bsnm{Appelö}, \binits{D.}},
\oauthor{\bsnm{Tennant}, \binits{D.M.}},
\oauthor{\bsnm{Petersson}, \binits{N.A.}},
\oauthor{\bsnm{Dubois}, \binits{J.L.}},
\oauthor{\bsnm{Rosen}, \binits{Y.J.}}:
Noise-specific beating in the higher-level {Ramsey} curves of a transmon qubit.
Applied Physics Letters
\textbf{122}(11)
(2023).
\doiurl{10.1063/5.0138811}
\end{botherref}
\endbibitem

\bibitem{amin_direct_2024}
\begin{botherref}
\oauthor{\bsnm{Amin}, \binits{K.R.}},
\oauthor{\bsnm{Eriksson}, \binits{A.M.}},
\oauthor{\bsnm{Kervinen}, \binits{M.}},
\oauthor{\bsnm{Andersson}, \binits{L.}},
\oauthor{\bsnm{Rehammar}, \binits{R.}},
\oauthor{\bsnm{Gasparinetti}, \binits{S.}}:
Direct detection of quasiparticle tunneling with a charge-sensitive superconducting sensor coupled to a waveguide
(2024)
{\href{https://arxiv.org/abs/2404.01277}{{arXiv:2404.01277}}}
{[quant-ph]}
\end{botherref}
\endbibitem

\bibitem{kamenov_suppression_2024}
\begin{botherref}
\oauthor{\bsnm{Kamenov}, \binits{P.}},
\oauthor{\bsnm{DiNapoli}, \binits{T.}},
\oauthor{\bsnm{Gershenson}, \binits{M.}},
\oauthor{\bsnm{Chakram}, \binits{S.}}:
Suppression of quasiparticle poisoning in transmon qubits by gap engineering
(2024)
{\href{https://arxiv.org/abs/2309.02655}{{arXiv:2309.02655}}}
{[quant-ph]}
\end{botherref}
\endbibitem

\bibitem{krause_quasiparticle_2024}
\begin{barticle}
\bauthor{\bsnm{Krause}, \binits{J.}},
\bauthor{\bsnm{Marchegiani}, \binits{G.}},
\bauthor{\bsnm{Janssen}, \binits{L.M.}},
\bauthor{\bsnm{Catelani}, \binits{G.}},
\bauthor{\bsnm{Ando}, \binits{Y.}},
\bauthor{\bsnm{Dickel}, \binits{C.}}:
\batitle{Quasiparticle effects in magnetic-field-resilient three-dimensional transmons}.
\bjtitle{Physical Review Applied}
\bvolume{22}(\bissue{4}),
\bfpage{044063}
(\byear{2024}).
\doiurl{10.1103/PhysRevApplied.22.044063}
\end{barticle}
\endbibitem

\bibitem{liu_observation_2024}
\begin{barticle}
\bauthor{\bsnm{Liu}, \binits{B.-J.}},
\bauthor{\bsnm{Wang}, \binits{Y.-Y.}},
\bauthor{\bsnm{Sheffer}, \binits{T.}},
\bauthor{\bsnm{Wang}, \binits{C.}}:
\batitle{Observation of {Discrete} {Charge} {States} of a {Coherent} {Two}-{Level} {System} in a {Superconducting} {Qubit}}.
\bjtitle{Physical Review Letters}
\bvolume{133}(\bissue{16}),
\bfpage{160602}
(\byear{2024}).
\doiurl{10.1103/PhysRevLett.133.160602}
\end{barticle}
\endbibitem

\bibitem{sundelin_real-time_2026}
\begin{botherref}
\oauthor{\bsnm{Sundelin}, \binits{S.}},
\oauthor{\bsnm{Andersson}, \binits{L.}},
\oauthor{\bsnm{Brunander}, \binits{H.}},
\oauthor{\bsnm{Gasparinetti}, \binits{S.}}:
Real-time detection of correlated quasiparticle tunneling events in a multi-qubit superconducting device
(2026)
{\href{https://arxiv.org/abs/2602.01945}{{arXiv:2602.01945}}}
{[quant-ph]}
\end{botherref}
\endbibitem

\bibitem{kerschbaum_assessing_2026}
\begin{botherref}
\oauthor{\bsnm{Kerschbaum}, \binits{M.}},
\oauthor{\bsnm{Wagner}, \binits{F.}},
\oauthor{\bsnm{Ognjanović}, \binits{U.}},
\oauthor{\bsnm{Vio}, \binits{G.}},
\oauthor{\bsnm{Knapp}, \binits{K.}},
\oauthor{\bsnm{Zanuz}, \binits{D.C.}},
\oauthor{\bsnm{Flasby}, \binits{A.}},
\oauthor{\bsnm{Panah}, \binits{M.B.}},
\oauthor{\bsnm{Wallraff}, \binits{A.}},
\oauthor{\bsnm{Besse}, \binits{J.-C.}}:
Assessing the {Sensitivity} of {Niobium}- and {Tantalum}-{Based} {Superconducting} {Qubits} to {Infrared} {Radiation}
(2026)
{\href{https://arxiv.org/abs/2602.05806}{{arXiv:2602.05806}}}
{[quant-ph]}
\end{botherref}
\endbibitem

\bibitem{Lledo2023}
\begin{barticle}
\bauthor{\bsnm{Lled{\'o}}, \binits{C.}},
\bauthor{\bsnm{Dassonneville}, \binits{R.}},
\bauthor{\bsnm{Moulinas}, \binits{A.}},
\bauthor{\bsnm{Cohen}, \binits{J.}},
\bauthor{\bsnm{Shillito}, \binits{R.}},
\bauthor{\bsnm{Bienfait}, \binits{A.}},
\bauthor{\bsnm{Huard}, \binits{B.}},
\bauthor{\bsnm{Blais}, \binits{A.}}:
\batitle{Cloaking a qubit in a cavity}.
\bjtitle{Nature Communications}
\bvolume{14}(\bissue{1}),
\bfpage{6313}
(\byear{2023}).
\doiurl{10.1038/s41467-023-42060-5}
\end{barticle}
\endbibitem

\bibitem{dassonneville_amplifying_2026}
\begin{barticle}
\bauthor{\bsnm{Dassonneville}, \binits{R.}},
\bauthor{\bsnm{Elouard}, \binits{C.}},
\bauthor{\bsnm{Cazali}, \binits{R.}},
\bauthor{\bsnm{Assouly}, \binits{R.}},
\bauthor{\bsnm{Bienfait}, \binits{A.}},
\bauthor{\bsnm{Auffèves}, \binits{A.}},
\bauthor{\bsnm{Huard}, \binits{B.}}:
\batitle{Amplifying microwave pulses with a single qubit engine fueled by quantum measurements}.
\bjtitle{Physical Review Research}
\bvolume{8}(\bissue{1}),
\bfpage{013228}
(\byear{2026}).
\doiurl{10.1103/rygc-bc3c}
\end{barticle}
\endbibitem

\bibitem{Chitta2022}
\begin{barticle}
\bauthor{\bsnm{Chitta}, \binits{S.P.}},
\bauthor{\bsnm{Zhao}, \binits{T.}},
\bauthor{\bsnm{Huang}, \binits{Z.}},
\bauthor{\bsnm{Mondragon-Shem}, \binits{I.}},
\bauthor{\bsnm{Koch}, \binits{J.}}:
\batitle{Computer-aided quantization and numerical analysis of superconducting circuits}.
\bjtitle{New Journal of Physics}
\bvolume{24}(\bissue{10}),
\bfpage{103020}
(\byear{2022}).
\doiurl{10.1088/1367-2630/ac94f2}
\end{barticle}
\endbibitem

\bibitem{Groszkowski2021}
\begin{barticle}
\bauthor{\bsnm{Groszkowski}, \binits{P.}},
\bauthor{\bsnm{Koch}, \binits{J.}}:
\batitle{Scqubits: a python package for superconducting qubits}.
\bjtitle{Quantum}
\bvolume{5},
\bfpage{583}
(\byear{2021}).
\doiurl{10.22331/q-2021-11-17-583}
\end{barticle}
\endbibitem

\bibitem{koch_charging_2009}
\begin{barticle}
\bauthor{\bsnm{Koch}, \binits{J.}},
\bauthor{\bsnm{Manucharyan}, \binits{V.}},
\bauthor{\bsnm{Devoret}, \binits{M.H.}},
\bauthor{\bsnm{Glazman}, \binits{L.I.}}:
\batitle{Charging {Effects} in the {Inductively} {Shunted} {Josephson} {Junction}}.
\bjtitle{Physical Review Letters}
\bvolume{103}(\bissue{21}),
\bfpage{217004}
(\byear{2009}).
\doiurl{10.1103/PhysRevLett.103.217004}
\end{barticle}
\endbibitem

\bibitem{khorramshahi_high-impedance_2025}
\begin{barticle}
\bauthor{\bsnm{Khorramshahi}, \binits{M.}},
\bauthor{\bsnm{Spiecker}, \binits{M.}},
\bauthor{\bsnm{Paluch}, \binits{P.}},
\bauthor{\bsnm{Geisert}, \binits{S.}},
\bauthor{\bsnm{Gosling}, \binits{N.}},
\bauthor{\bsnm{Zapata}, \binits{N.}},
\bauthor{\bsnm{Brauch}, \binits{L.}},
\bauthor{\bsnm{K\"ubel}, \binits{C.}},
\bauthor{\bsnm{Dehm}, \binits{S.}},
\bauthor{\bsnm{Krupke}, \binits{R.}},
\bauthor{\bsnm{Wernsdorfer}, \binits{W.}},
\bauthor{\bsnm{Pop}, \binits{I.M.}},
\bauthor{\bsnm{Reisinger}, \binits{T.}}:
\batitle{High-impedance granular-aluminum ring resonators}.
\bjtitle{Phys. Rev. Appl.}
\bvolume{24},
\bfpage{024066}
(\byear{2025}).
\doiurl{10.1103/n73t-7q1n}
\end{barticle}
\endbibitem

\bibitem{pechenezhskiy_superconducting_2020}
\begin{barticle}
\bauthor{\bsnm{Pechenezhskiy}, \binits{I.V.}},
\bauthor{\bsnm{Mencia}, \binits{R.A.}},
\bauthor{\bsnm{Nguyen}, \binits{L.B.}},
\bauthor{\bsnm{Lin}, \binits{Y.-H.}},
\bauthor{\bsnm{Manucharyan}, \binits{V.E.}}:
\batitle{The superconducting quasicharge qubit}.
\bjtitle{Nature}
\bvolume{585}(\bissue{7825}),
\bfpage{368}--\blpage{371}
(\byear{2020}).
\doiurl{10.1038/s41586-020-2687-9}
\end{barticle}
\endbibitem

\bibitem{junger_implementation_2025}
\begin{barticle}
\bauthor{\bsnm{Jünger}, \binits{C.}},
\bauthor{\bsnm{Chistolini}, \binits{T.}},
\bauthor{\bsnm{Nguyen}, \binits{L.B.}},
\bauthor{\bsnm{Kim}, \binits{H.}},
\bauthor{\bsnm{Chen}, \binits{L.}},
\bauthor{\bsnm{Ersevim}, \binits{T.}},
\bauthor{\bsnm{Livingston}, \binits{W.}},
\bauthor{\bsnm{Koolstra}, \binits{G.}},
\bauthor{\bsnm{Santiago}, \binits{D.I.}},
\bauthor{\bsnm{Siddiqi}, \binits{I.}}:
\batitle{Implementation of scalable suspended superinductors}.
\bjtitle{Applied Physics Letters}
\bvolume{126}(\bissue{4}),
\bfpage{044003}
(\byear{2025}).
\doiurl{10.1063/5.0250341}
\end{barticle}
\endbibitem

\bibitem{mcfadden_interface-sensitive_2025}
\begin{barticle}
\bauthor{\bsnm{McFadden}, \binits{A.P.}},
\bauthor{\bsnm{Larson}, \binits{T.F.Q.}},
\bauthor{\bsnm{Gill}, \binits{S.}},
\bauthor{\bsnm{Dixit}, \binits{A.V.}},
\bauthor{\bsnm{Simmonds}, \binits{R.}},
\bauthor{\bsnm{Lecocq}, \binits{F.}},
\bauthor{\bsnm{Oh}, \binits{J.}},
\bauthor{\bsnm{Zhou}, \binits{L.}}:
\batitle{Interface-sensitive microwave loss in superconducting tantalum films sputtered on c-plane sapphire}.
\bjtitle{Physical Review Materials}
\bvolume{9}(\bissue{9}),
\bfpage{096201}
(\byear{2025}).
\doiurl{10.1103/lwn1-fznb}
\end{barticle}
\endbibitem

\bibitem{anbalagan_revealing_2025}
\begin{barticle}
\bauthor{\bsnm{Anbalagan}, \binits{A.k.}},
\bauthor{\bsnm{Cummings}, \binits{R.}},
\bauthor{\bsnm{Zhou}, \binits{C.}},
\bauthor{\bsnm{Mun}, \binits{J.}},
\bauthor{\bsnm{Stanic}, \binits{V.}},
\bauthor{\bsnm{Jordan-Sweet}, \binits{J.}},
\bauthor{\bsnm{Yao}, \binits{J.}},
\bauthor{\bsnm{Kisslinger}, \binits{K.}},
\bauthor{\bsnm{Weiland}, \binits{C.}},
\bauthor{\bsnm{Nykypanchuk}, \binits{D.}},
\bauthor{\bsnm{Hulbert}, \binits{S.L.}},
\bauthor{\bsnm{Li}, \binits{Q.}},
\bauthor{\bsnm{Zhu}, \binits{Y.}},
\bauthor{\bsnm{Liu}, \binits{M.}},
\bauthor{\bsnm{Sushko}, \binits{P.V.}},
\bauthor{\bsnm{Walter}, \binits{A.L.}},
\bauthor{\bsnm{Barbour}, \binits{A.M.}}:
\batitle{Revealing the {Origin} and {Nature} of the {Buried} {Metal}-{Substrate} {Interface} {Layer} in {Ta}/{Sapphire} {Superconducting} {Films}}.
\bjtitle{Advanced Science}
\bvolume{12}(\bissue{17}),
\bfpage{2413058}
(\byear{2025}).
\doiurl{10.1002/advs.202413058}
\end{barticle}
\endbibitem

\bibitem{olszewski_krypton-sputtered_2026}
\begin{botherref}
\oauthor{\bsnm{Olszewski}, \binits{M.W.}},
\oauthor{\bsnm{Kong}, \binits{L.}},
\oauthor{\bsnm{Reinhardt}, \binits{S.}},
\oauthor{\bsnm{Tong}, \binits{D.}},
\oauthor{\bsnm{Du}, \binits{X.}},
\oauthor{\bsnm{Gianluca}, \binits{G.D.}},
\oauthor{\bsnm{Lu}, \binits{H.}},
\oauthor{\bsnm{Roy}, \binits{S.}},
\oauthor{\bsnm{Zhang}, \binits{L.}},
\oauthor{\bsnm{Biedron}, \binits{A.B.}},
\oauthor{\bsnm{Muller}, \binits{D.A.}},
\oauthor{\bsnm{Fatemi}, \binits{V.}}:
Krypton-sputtered tantalum films for scalable high-performance quantum devices
(2026)
{\href{https://arxiv.org/abs/2601.20091}{{arXiv:2601.20091}}}
{[quant-ph]}
\end{botherref}
\endbibitem

\bibitem{Nho2025}
\begin{barticle}
\bauthor{\bsnm{Nho}, \binits{H.}},
\bauthor{\bsnm{Connolly}, \binits{T.}},
\bauthor{\bsnm{Kurilovich}, \binits{P.D.}},
\bauthor{\bsnm{Diamond}, \binits{S.}},
\bauthor{\bsnm{B\o{}ttcher}, \binits{C.G.L.}},
\bauthor{\bsnm{Glazman}, \binits{L.I.}},
\bauthor{\bsnm{Devoret}, \binits{M.H.}}:
\batitle{Recovery dynamics of a gap-engineered transmon after a quasiparticle burst}.
\bjtitle{Phys. Rev. Lett.}
\bvolume{136},
\bfpage{050601}
(\byear{2026}).
\doiurl{10.1103/ql6q-wfpn}
\end{barticle}
\endbibitem

\bibitem{Kurilovich2025}
\begin{botherref}
\oauthor{\bsnm{Kurilovich}, \binits{V.D.}},
\oauthor{\bsnm{Roberts}, \binits{G.}},
\oauthor{\bsnm{Martin}, \binits{L.S.}},
\oauthor{\bsnm{McEwen}, \binits{M.}},
\oauthor{\bsnm{Eickbusch}, \binits{A.}},
\oauthor{\bsnm{Faoro}, \binits{L.}},
\oauthor{\bsnm{Ioffe}, \binits{L.B.}},
\oauthor{\bsnm{Atalaya}, \binits{J.}},
\oauthor{\bsnm{Bilmes}, \binits{A.}},
\oauthor{\bsnm{Kreikebaum}, \binits{J.M.}},
\oauthor{\bsnm{Bengtsson}, \binits{A.}},
\oauthor{\bsnm{Klimov}, \binits{P.}},
\oauthor{\bsnm{Neeley}, \binits{M.}},
\oauthor{\bsnm{Mruczkiewicz}, \binits{W.}},
\oauthor{\bsnm{Miao}, \binits{K.}},
\oauthor{\bsnm{Aleiner}, \binits{I.L.}},
\oauthor{\bsnm{Kelly}, \binits{J.}},
\oauthor{\bsnm{Chen}, \binits{Y.}},
\oauthor{\bsnm{Satzinger}, \binits{K.}},
\oauthor{\bsnm{Opremcak}, \binits{A.}}:
Correlated error bursts in a gap-engineered superconducting qubit array
(2025)
{\href{https://arxiv.org/abs/2506.18228}{{arXiv:2506.18228}}}
{[quant-ph]}
\end{botherref}
\endbibitem

\bibitem{coatings11101206}
\begin{botherref}
\oauthor{\bsnm{Fedorov}, \binits{P.}},
\oauthor{\bsnm{Nazarov}, \binits{D.}},
\oauthor{\bsnm{Medvedev}, \binits{O.}},
\oauthor{\bsnm{Koshtyal}, \binits{Y.}},
\oauthor{\bsnm{Rumyantsev}, \binits{A.}},
\oauthor{\bsnm{Tolmachev}, \binits{V.}},
\oauthor{\bsnm{Popovich}, \binits{A.}},
\oauthor{\bsnm{Maximov}, \binits{M.Y.}}:
Plasma enhanced atomic layer deposition of tantalum (v) oxide.
Coatings
\textbf{11}(10)
(2021).
\doiurl{10.3390/coatings11101206}
\end{botherref}
\endbibitem

\bibitem{assouly:tel-04057646}
\begin{botherref}
\oauthor{\bsnm{Assouly}, \binits{R.}}:
{Superconducting Quantum Node for Quantum Sensing}.
Theses,
{Ecole normale sup{\'e}rieure de lyon - ENS LYON}
(December 2022).
\url{https://theses.hal.science/tel-04057646}
\end{botherref}
\endbibitem

\bibitem{Macklin2015}
\begin{barticle}
\bauthor{\bsnm{Macklin}, \binits{C.}},
\bauthor{\bsnm{O’Brien}, \binits{K.}},
\bauthor{\bsnm{Hover}, \binits{D.}},
\bauthor{\bsnm{Schwartz}, \binits{M.E.}},
\bauthor{\bsnm{Bolkhovsky}, \binits{V.}},
\bauthor{\bsnm{Zhang}, \binits{X.}},
\bauthor{\bsnm{Oliver}, \binits{W.D.}},
\bauthor{\bsnm{Siddiqi}, \binits{I.}}:
\batitle{A near–quantum-limited {Josephson} traveling-wave parametric amplifier}.
\bjtitle{Science}
\bvolume{350}(\bissue{6258}),
\bfpage{307}--\blpage{310}
(\byear{2015}).
\doiurl{10.1126/science.aaa8525}
\end{barticle}
\endbibitem

\bibitem{willsch_observation_2024}
\begin{barticle}
\bauthor{\bsnm{Willsch}, \binits{D.}},
\bauthor{\bsnm{Rieger}, \binits{D.}},
\bauthor{\bsnm{Winkel}, \binits{P.}},
\bauthor{\bsnm{Willsch}, \binits{M.}},
\bauthor{\bsnm{Dickel}, \binits{C.}},
\bauthor{\bsnm{Krause}, \binits{J.}},
\bauthor{\bsnm{Ando}, \binits{Y.}},
\bauthor{\bsnm{Lescanne}, \binits{R.}},
\bauthor{\bsnm{Leghtas}, \binits{Z.}},
\bauthor{\bsnm{Bronn}, \binits{N.T.}},
\bauthor{\bsnm{Deb}, \binits{P.}},
\bauthor{\bsnm{Lanes}, \binits{O.}},
\bauthor{\bsnm{Minev}, \binits{Z.K.}},
\bauthor{\bsnm{Dennig}, \binits{B.}},
\bauthor{\bsnm{Geisert}, \binits{S.}},
\bauthor{\bsnm{Günzler}, \binits{S.}},
\bauthor{\bsnm{Ihssen}, \binits{S.}},
\bauthor{\bsnm{Paluch}, \binits{P.}},
\bauthor{\bsnm{Reisinger}, \binits{T.}},
\bauthor{\bsnm{Hanna}, \binits{R.}},
\bauthor{\bsnm{Bae}, \binits{J.H.}},
\bauthor{\bsnm{Schüffelgen}, \binits{P.}},
\bauthor{\bsnm{Grützmacher}, \binits{D.}},
\bauthor{\bsnm{Buimaga-Iarinca}, \binits{L.}},
\bauthor{\bsnm{Morari}, \binits{C.}},
\bauthor{\bsnm{Wernsdorfer}, \binits{W.}},
\bauthor{\bsnm{DiVincenzo}, \binits{D.P.}},
\bauthor{\bsnm{Michielsen}, \binits{K.}},
\bauthor{\bsnm{Catelani}, \binits{G.}},
\bauthor{\bsnm{Pop}, \binits{I.M.}}:
\batitle{Observation of {Josephson} harmonics in tunnel junctions}.
\bjtitle{Nature Physics}
\bvolume{20}(\bissue{5}),
\bfpage{815}--\blpage{821}
(\byear{2024}).
\doiurl{10.1038/s41567-024-02400-8}
\end{barticle}
\endbibitem

\bibitem{kim_emergent_2025}
\begin{botherref}
\oauthor{\bsnm{Kim}, \binits{J.}},
\oauthor{\bsnm{Hays}, \binits{M.}},
\oauthor{\bsnm{Rosen}, \binits{I.T.}},
\oauthor{\bsnm{An}, \binits{J.}},
\oauthor{\bsnm{Zhang}, \binits{H.}},
\oauthor{\bsnm{Goswami}, \binits{A.}},
\oauthor{\bsnm{Azar}, \binits{K.}},
\oauthor{\bsnm{Gertler}, \binits{J.M.}},
\oauthor{\bsnm{Niedzielski}, \binits{B.M.}},
\oauthor{\bsnm{Schwartz}, \binits{M.E.}},
\oauthor{\bsnm{Orlando}, \binits{T.P.}},
\oauthor{\bsnm{Grover}, \binits{J.A.}},
\oauthor{\bsnm{Serniak}, \binits{K.}},
\oauthor{\bsnm{Oliver}, \binits{W.D.}}:
Emergent {Harmonics} in {Josephson} {Tunnel} {Junctions} {Due} to {Series} {Inductance}
(2025)
{\href{https://arxiv.org/abs/2507.08171}{{arXiv:2507.08171}}}
{[quant-ph]}
\end{botherref}
\endbibitem

\bibitem{wang_high-e_je_c_2025}
\begin{barticle}
\bauthor{\bsnm{Wang}, \binits{Z.}},
\bauthor{\bsnm{Parker}, \binits{R.W.}},
\bauthor{\bsnm{Champion}, \binits{E.}},
\bauthor{\bsnm{Blok}, \binits{M.S.}}:
\batitle{High-{$E_J/E_C$} transmon qudits with up to 12 levels}.
\bjtitle{Physical Review Applied}
\bvolume{23}(\bissue{3}),
\bfpage{034046}
(\byear{2025}).
\doiurl{10.1103/PhysRevApplied.23.034046}
\end{barticle}
\endbibitem

\bibitem{shagalov_higher_2025}
\begin{botherref}
\oauthor{\bsnm{Shagalov}, \binits{K.}},
\oauthor{\bsnm{Feldstein-Bofill}, \binits{D.}},
\oauthor{\bsnm{Jakobsen}, \binits{L.U.}},
\oauthor{\bsnm{Sun}, \binits{Z.}},
\oauthor{\bsnm{Wied}, \binits{C.}},
\oauthor{\bsnm{Paulsen}, \binits{A.T.J.}},
\oauthor{\bsnm{Severin}, \binits{J.B.}},
\oauthor{\bsnm{Marciniak}, \binits{M.A.}},
\oauthor{\bsnm{Potts}, \binits{C.A.}},
\oauthor{\bsnm{Kringhøj}, \binits{A.}},
\oauthor{\bsnm{Hastrup}, \binits{J.}},
\oauthor{\bsnm{Flensberg}, \binits{K.}},
\oauthor{\bsnm{Krøjer}, \binits{S.}},
\oauthor{\bsnm{Kjaergaard}, \binits{M.}}:
Higher {Josephson} harmonics in a tunable double-junction transmon qubit
(2025)
{\href{https://arxiv.org/abs/2512.08470}{{arXiv:2512.08470}}}
{[quant-ph]}
\end{botherref}
\endbibitem

\bibitem{xia_exceeding_2025}
\begin{botherref}
\oauthor{\bsnm{Xia}, \binits{M.}},
\oauthor{\bsnm{Lledó}, \binits{C.}},
\oauthor{\bsnm{Capocci}, \binits{M.}},
\oauthor{\bsnm{Repicky}, \binits{J.}},
\oauthor{\bsnm{D'Anjou}, \binits{B.}},
\oauthor{\bsnm{Mondragon-Shem}, \binits{I.}},
\oauthor{\bsnm{Kaufman}, \binits{R.}},
\oauthor{\bsnm{Koch}, \binits{J.}},
\oauthor{\bsnm{Blais}, \binits{A.}},
\oauthor{\bsnm{Hatridge}, \binits{M.}}:
Exceeding the {Parametric} {Drive} {Strength} {Threshold} in {Nonlinear} {Circuits}
(2025)
{\href{https://arxiv.org/abs/2506.03456}{{arXiv:2506.03456}}}
{[quant-ph]}
\end{botherref}
\endbibitem

\end{thebibliography}
\end{document}